\documentclass[twoside,preprintnumbers,amsmath,amssymb,pacs,shownopacs,nofootinbib,notitlepage]{revtex4-1}

\usepackage{graphicx,amsmath,amsfonts,amssymb,dcolumn,amsthm,euscript,braket}
\usepackage{slashed}
\usepackage{xcolor}
\usepackage[symbol*]{footmisc}
\usepackage{caption}
\usepackage{subcaption}
\usepackage[margin=10pt,font=small,labelfont=bf,
labelsep=endash, format=hang, justification=raggedright]{caption}
\usepackage{float}
\usepackage[makeroom]{cancel}

\begin{document}

\title{{\bf Landau-Khalatnikov-Fradkin Transformations, Nielsen Identities, Their Equivalence and Implications for QCD}}

\author{T.~De Meerleer$^{\dag}$, D.~Dudal$^{\dag,\ddag}$, S.~P.~Sorella$^\S$}
\email{tim.demeerleer@kuleuven.be, david.dudal@kuleuven.be, silvio.sorella@gmail.com}
\affiliation{$\dag$ KU Leuven Campus Kortrijk--Kulak, Department of Physics, Etienne Sabbelaan 53 bus 7657, 8500 Kortrijk, Belgium\\ $\ddag$ Ghent University, Department of Physics and Astronomy, Krijgslaan 281-S9, 9000 Gent, Belgium \\$^\S$ Instituto de F\'isica Te\'orica, Rua S\~ao Francisco Xavier 524, 20550-013, Maracan\~a, Rio de Janeiro, Brasil}

\author{P.~Dall'Olio$^{\ast}$, A.~Bashir$^{\ast \ast}$}
\email{adnan.bashir@umich.mx, pietro@matmor.unam.mx}
\affiliation{$\ast$Centro de Ciencias Matem\'{a}ticas, Unam - Campus Morelia, Antigua Carretera a P\'atzcuaro 8701
Col. Ex Hacienda San Jos\'e de la Huerta
Morelia, Michoac\'an  58089, M\'exico \\
$\ast \ast$ Instituto de F\'{i}sica y Matem\'{a}ticas, Universidad Michoacana de San Nicol\'{a}s de Hidalgo,
Edificio C-3, Ciudad Universitaria, Morelia, Michoac\'{a}n 58040, M\'{e}xico}

\begin{abstract}
The Landau-Khalatnikov-Fradkin transformations (LKFTs) represent an important tool for probing the gauge dependence of the correlation functions within the class of linear covariant gauges. Recently these transformations have been derived from first principles in the context of non-Abelian gauge theory (QCD) introducing a gauge invariant transverse gauge field expressible as an infinite power series in a Stueckelberg field. In this work we explicitly calculate the transformation for the gluon propagator, reproducing its dependence on the gauge parameter at the one loop level and elucidating the role of the extra fields involved in this theoretical framework. Later on, employing a unifying scheme based upon the BRST symmetry and a resulting generalized Slavnov-Taylor identity, we establish the equivalence between the LKFTs and the Nielsen identities which are also known to connect results in different gauges.
\end{abstract}

\maketitle

\section{Introduction}
Gauge symmetries are an ubiquitous feature in our theoretical understanding of physical interactions at their fundamental level. The quantum field theory of the Standard Model of elementary particles as well as the classical theory of General Relativity describing gravitational force contain redundant degrees of freedom in their dynamical field content. They transform non-trivially under a local gauge transformation which leaves the action invariant.

Unlike global (rigid) symmetries, gauge symmetries do not rotate physical observables which are manifestly gauge invariant quantities. In a continuum formulation of the theory, these quantities are generally extracted from gauge dependent correlation (or Green) functions. It is a non-trivial problem to understand how the gauge dependence gets cancelled in calculating physical observables both in perturbative and non perturbative calculations.

In the continuum treatment of field theories, which preserves Lorentz symmetry (or its Euclidean counterpart), one can
work with a gauge fixing procedure \`a la Faddeev-Popov \cite{Faddeev:1967}. Within the linear covariant gauges considered here, this
procedure ensures that the gauge dependence of the correlation functions is manifest in the appearance of the gauge fixing parameter in their explicit expressions.

LKFTs \cite{Landau:1955, Fradkin:1955} (see also \cite{Johnson:1959, Zumino:1960} for a derivation using functional methods) are a set of identities which interpolate an arbitrary $n$-point correlation function for different values of the gauge parameter. These transformations have been investigated primarily in the Abelian case (QED)
\cite{Burden:1993, Bashir:1999, Bashir:2000, Bashir:2002, Delbourgo:2004, Raya:2005, Bashir:2007, Bashir:2008, Bashir:2009, Ahmadiniaz:2016, Concha:2016, Jia:2017pl, Jia:2017pr, Dallolio:2019, Kotikov:2019} (see however \cite{Aslam:2016}) and have focused exclusively on the fermion propagator.

In the non-Abelian case (QCD) the constraints imposed by the gauge transformations  on the correlation functions have been mainly studied in the form of the Slavnov-Taylor identities \cite{Slavnov:1972} (their Abelian counterpart are the Ward-Takahashi identities \cite{Ward:1950}). These identities can be derived by exploiting the invariance of the effective action under the BRST symmetry transformation \cite{Becchi:1976} which guarantees the renormalizabilty of the theory. Besides being an essential tool in order to prove renormalizability to all orders, these identities have constituted the main guide to constructing viable \emph{ans\"{a}tze} for the dressed fermion-boson vertex inside the fermion gap equation \cite{Curtis:1990, Bashir:1994, Delbourgo:2007, Kizilersu:2009, Bashir:2011, Aguilar:2011, Raya:2011, Bashir:2012, Qin:2013,
Kizilersu:2015, Albino:2016, Aguilar:2017, Albino:2017, Albino:2019}. The latter is a non-perturbative key element in the continuum for studying quark confinement and dynamical chiral symmetry breaking (DCSB) \cite{Roberts:1994, Alkofer:2001}. The gap equation for the quark propagator constitutes one piece of an infinite tower of non-linear coupled \emph{Schwinger-Dyson} equations for the $n$-point 1PI correlation functions. This system of equations needs to be truncated at some level in order for it to be analytically and numerically tractable. Therefore, there is an intrinsic need to provide a model for the higher order Green functions which are left out unsolved on employing a truncation.

The gauge covariance of the correlation functions, expressed by the LKFTs, should further restrain the possible form of the fermion-boson vertex. In addition to be consistent with its own LKFT, any choice of the vertex should yield the correct gauge covariance of the quark propagator within the gap equation, and also guarantee the gauge invariance of physical observables.

In \cite{DeMeerleer:2018txc}, a derivation of the LKFTs for the $n$-point correlation functions has been detailed, employing gauge invariant composite operators $A^h_\mu$ and $\psi^h$ which involve a Stueckelberg type scalar field. These composite fields, originally introduced  in an attempt to construct gauge invariant colored states \cite{Lavelle:1995ty}, have recently received renewed spotlight in the context of gauge-fixing procedure at a non-perturbative level. The context is the problem of \emph{Gribov copies} \cite{Gribov:1978}, by extending the Gribov-Zwanziger scenario, originally formulated in Landau gauge, to the class of linear covariant gauges, while preserving a nilpotent BRST symmetry \cite{Capri:2015brs, Capri:2016lin, Capri:2016grib}.
A derivation of the LKFTs through the introduction of a Stueckelberg field had already been carried out in QED \cite{Sonoda:2001}. The formalism developed in \cite{DeMeerleer:2018txc} generalizes this approach to the non-Abelian case.

In the following sections, we set out to verify the perturbative validity of the latter formalism which involves the appearance of new dynamical fields. We derive the LKFT for the gluon propagator and from there evaluate its expression at the one loop level. Our LKFT expression confirms that the longitudinal part of the gauge boson propagator does not receive any quantum corrections, a result which is also true for the Abelian case.
However, unlike the case of a massless U(1) gauge boson, the transverse part contains a non-trivial dependence on the gauge parameter.
Our LKFT-based result correctly reproduces this dependence. Furthermore, using the extended source formalism, we also discuss and establish the equivalence between the LKFTs and the Nielsen identities.

\section{Classical action}
We consider the following classical action in Euclidean space \cite{Capri:2016a2, Capri:2017ren, DeMeerleer:2018txc}
\begin{equation} \label{action}
S= S_{QCD} + S_{gf} + S_h,
\end{equation}
where $S_{QCD}$ is the usual gauge invariant QCD action which encodes the dynamics of quarks and gluons
\begin{equation}
S_{QCD} = \int d^D x \left[\frac{1}{4} F^a_{\mu \nu}F^a_{\mu \nu} + \bar \psi \left(\gamma_\mu D_\mu+m_f \right)\psi \right],
\end{equation}
where only one flavour $f$ of quarks is considered for the sake of simplicity. $F^a_{\mu \nu} = \partial_\mu A^a_\nu - \partial_\nu A^a_\mu + g f^{abc}A^b_\mu A^c_\nu$ is the gluon field strength tensor and $D_\mu = \partial_\mu - i g T^a A^a_\mu$ is the covariant derivative, $T^a$ being the generators of the SU($N$) group which satisfy the Lie algebra $[T^a, T^b]=if^{abc}T^c$.  $S_{gf}$ includes the gauge-fixing terms in linear covariant gauges
\begin{equation} \label{eq:gf}
S_{gf} = \int d^D x \left[ \frac{\alpha}{2}b^a b^a + i b^a \partial_\mu A^a_\mu + \bar c^a \partial_\mu D^{ab}_\mu c^b\right],
\end{equation}
where $b^a$ is the Nakanishi-Lautrup auxiliary field which implements the gauge condition  \cite{Nakanishi:1966}, $\alpha$ is the gauge parameter, $c^a(x)$ and $\bar c^a(x)$ are the anti-commuting ghost fields which yield the exponential representation of the Jacobian arising in the gauge fixing procedure, and $D^{ab}_\mu=\partial_\mu \delta^{ab}-gf^{abc}A^c_\mu$ is the covariant derivative in the adjoint representation. The \emph{Landau} gauge corresponds to $\alpha = 0$, for which the auxiliary field $b^a$ strictly enforces the transversality condition $\partial_\mu A^a_\mu=0$.

This standard gauge-fixed QCD action is augmented by the term
\begin{equation} \label{S_h}
S_h = \int d^D x \left[\tau^a \partial_\mu A^{h,a}_\mu +\bar \eta^a\, \partial_\mu D^{ab}_\mu(A^h) \eta^b\right],
\end{equation}
where $A_{\mu}^h$ is the composite operator which incorporates the gauge field $A_\mu$ and the scalar Stueckelberg field $\xi$,
\begin{equation} \label{Ahloc}
A_\mu^h = h^\dagger A_\mu h+\frac{i}{g}h^\dagger \partial_\mu h, \qquad h=e^{igT^a \xi^a}.
\end{equation}
Its gauge invariance is guaranteed by the action of the SU($N$) gauge transformation $U=e^{igT^a \alpha^a}$ on $h$
\begin{equation}
\begin{split}
& h^U\equiv U^\dagger h\,, \qquad A^U_\mu = U^\dagger A_\mu U + \frac{i}{g} U^\dagger \partial_\mu U \\
&\Longrightarrow (A^h_\mu)^U = (h^\dagger)^U A_{\mu}^U h^U + \frac{i}{g}(h^\dagger)^U \partial_\mu h^U = A^h_\mu.
\end{split}
\end{equation}
$A^h_\mu$ is forced to be transverse through the introduction of the auxiliary field $\tau^a$ in $S_h$. It is the localized representation of the gauge invariant non-local operator that minimizes the functional $\int d^D x\, \mathrm{Tr}\, A^U_\mu A^U_\mu$ along a gauge orbit, given by (see for instance \cite{Capri:2015brs})
\begin{equation} \label{Ahnonloc}
A^h_\mu = \left(\delta_{\mu \nu} - \frac{\partial_\mu \partial_\nu}{\partial^2}\right) \left(A_\nu -ig\left[\frac{1}{\partial^2}\partial_\sigma A_\sigma, A_\nu \right] +\frac{ig}{2} \left[\frac{1}{\partial^2}\partial_\sigma A_\sigma, \partial_\nu \frac{1}{\partial^2}\partial_\sigma A_\sigma  \right] + {\cal O}(A^3)\right).
\end{equation}
Each term in this non-local expression contains at least one factor of the gauge field divergence, which makes it explicit how in Landau gauge ($\partial_\mu A_\mu=0$) $A^h_\mu = A_\mu$.
By expanding \eqref{Ahloc} in powers of $\xi$ and imposing the transversality condition, one can iteratively solve for the Stueckelberg field  $\xi$,
\begin{equation} \label{xieom}
\xi= \frac{1}{\partial^2}\partial_\mu A_\mu+ i\frac{g}{\partial^2}\left[\partial_\mu A_\mu,\frac{\partial_\sigma}{\partial^2}A_\sigma\right]+i\frac{g}{\partial^2}\left[A_\mu,\frac{\partial_\mu}{\partial^2}\partial_\sigma A_\sigma\right]+i\frac{g}{2\partial^2}\left[\frac{\partial_\mu}{\partial^2}A_\mu,\partial_\sigma A_\sigma\right]+ {\cal O}(A^3).
\end{equation}
and recover the non-local representation given in \eqref{Ahnonloc}. The introduction of a new pair of Grassmannian ghost fields $\eta$ and $\bar \eta$ in \eqref{S_h} is required in order to take care of the non-trivial Jacobian arising through the dependence of $A^h_\mu$ on the Stuckelberg field. In fact, if one integrates back the auxiliary field $\tau$ and the pair of ghost fields, one is left with the path integration over the Stueckelberg field, yielding
\begin{equation}
\int D\xi \, \delta \! \left(\partial_\mu A^h_\mu\right) \det \left(-\partial_\mu D_\mu( A^h_\mu )\right) = 1.
\end{equation}
We point out that this result is based on the perturbative solution\footnote{We thank Urko Reinosa for discussion on this point.} of the $\tau$-equation of motion, which results in the \emph{unique perturbative} series solution \eqref{xieom}. Beyond perturbation theory, this uniqueness might not prevail. This is the same identity introduced in the Faddeev-Popov gauge-fixing procedure. This means that the insertion of $S_h$ does not change the physical content of the theory. It leads to a classical action that is non-polynomial, given the infinite number of terms obtained by expanding \eqref{Ahloc} in terms of $\xi$
\begin{equation} \label{Ahexp}
A^{h,a}_\mu = A^a_\mu - D^{ab}_\mu \xi^b -\frac{g}{2} f^{abc} \xi^b D^{cd}_\mu \xi^d -\frac{g^2}{3!} f^{abd}f^{dce} \xi^b \xi^c D^{ef}_\mu \xi^f + {\cal O}(\xi^4).
\end{equation}
Despite this fact, we emphasize that the Lagrangian is still local since each term in \eqref{Ahexp} is comprised of at most one derivative of $\xi$. This feature  is easily understood through dimensional analysis, the Stueckelberg field being dimensionless. This scenario shares similarities with the one encountered in the non-linear sigma models.

Due to the gauge invariance of $A^h_\mu$, the new term represented by $S_h$ clearly does not spoil the standard BRST symmetry of the classical action; while the new fields $\tau$, $\eta$ and $\bar \eta$ are BRST singlets, the BRST transformation of the Stueckelberg field can be obtained iteratively from the transformation of $h$, ($sh^{ij}=-ig c^a (T^a)^{ik} h^{kj}$), yielding
\begin{equation}
s \xi^a = -c^a + \frac{g}{2}f^{abc}c^b\xi^c-\frac{g^2}{12}f^{abc}f^{bde}c^d\xi^e\xi^c + {\cal O}(\xi^3).
\end{equation}
The introduction of the gauge invariant field $A^h_\mu$ makes it straightforward to accommodate a dimension $d=2$ gluon operator in a gauge invariant fashion. It is achieved by adding the term $\frac{1}{2}m^2 A^{h,a}_\mu A^{h,a}_\mu$ to the action. This term takes into account the non-perturbative formation of a dimension two gluon condensate \cite{Chetyrkin:1999, Gubarev:2001as, Gubarev:2001, Boucaud:2001, Verschelde:2001} which is responsible for the infrared saturation of the gluon propagator, a behavior emerged in different continuum approaches \cite{Cornwall:1982, Aguilar:2004,Aguilar:2008xm, Dudal:2008,Ayala:2012pb,Siringo:2015wtx,Cyrol:2017ewj} and observed in lattice simulations \cite{Cucchieri:2007, Cucchieri:2008, Bogolubsky:2007, Oliveira:2012eh}. In particular, in Landau gauge, where $A^h_\mu= A_\mu$, this gauge invariant massive term will reduce to the massive extension of Yang-Mills theory, known as Curci-Ferrari model, which has recently been proved very successful in reproducing lattice results for the correlation functions \cite{Tissier:2011,Gracey:2019xom}.
Since we are merely interested in reproducing the correct gauge covariance of the gluon propagator at a perturbative level for now, we will not dwell on the effects of this massive operator.

It is noteworthy that, whether the massive operator is included or not, action \eqref{action} defines a quantum theory which is renormalizable to all orders in perturbation theory \cite{Capri:2016a2, Capri:2017ren, Capri:2018ir}. It is quite remarkable given the infinite number of interactions present. For this fundamental property to hold, the transversality of the field $A^h_\mu$ is the key factor which distinguishes this formulation from the original non-power counting renormalizable Stueckelberg action \cite{Ruegg:2004}.

\section{Landau-Khalatnikov-Fradkin transformations}
Although the addition of $S_h$ to the classical action does not change the physical content of the theory and appears to present a pointless complication, the introduction of the gauge invariant field $A^h_\mu$ makes it straightforward to derive the gauge covariance of the correlation functions (see \cite{DeMeerleer:2018txc} for an alternative derivation based on functional integration). In fact, we can define the corresponding gauge invariant operators for the fermion fields
\begin{equation} \label{psih}
\begin{split}
&\psi^h = h^\dagger \psi, \qquad \bar \psi^h = \bar \psi h, \\
\Longrightarrow &(\psi^h)^U = (h^\dagger)^U \psi^U=h^\dagger U U^\dagger \psi = \psi^h, \quad (\bar \psi^h)^U = \bar \psi^h,
\end{split}
\end{equation}
which can be expanded in terms of the Stuckelberg field $\xi$ as
\begin{equation}
(\psi^h)^{i} =  \psi^i -ig \xi^{a}(T^a)^{ij} \psi^j - \frac{g^2}{2}\xi^a \xi^b (T^a)^{ik}(T^b)^{kj} \psi^j + {\cal O}(\xi^3). \label{psih_expansion}
\end{equation}
Combining these gauge invariant fields, one can derive the LKFTs for the correlation function of an arbitrary product of gauge and fermion fields. The crucial point is that the correlation function of a product of gauge invariant (and therefore BRST invariant) operators $A^h_\mu$ and $\psi^h$ does not depend on the gauge parameter \cite{Capri:2016a2, Capri:2017ren, Capri:2018un}. Therefore, we have
\begin{equation} \label{gaugeind}
\resizebox{.95\hsize}{!}
{$\langle  A_{\mu_1}^{h}(x_1) \dots A_{\mu_n}^{h}(x_n) \bar \psi^{h} (y_1) \psi^{h}(z_1) \dots \bar\psi^{h} (y_m) \psi^{h}(z_m)  \rangle_{\alpha}  =
\langle  A_{\mu_1}^{h}(x_1) \dots A_{\mu_n}^{h}(x_n) \bar \psi^{h} (y_1) \psi^{h}(z_1) \dots \bar\psi^{h} (y_m) \psi^{h}(z_m)  \rangle_{\alpha'}$},
\end{equation}
where the subscript $\alpha$ ($\alpha'$) refers to the value of the gauge parameter used to evaluate the correlation function. If one now sets $\alpha'=0$ (Landau gauge), the rhs of \eqref{gaugeind} reduces to the correlation function under interest of the usual elementary fields, since in this particular gauge the Stueckelberg field $\xi$ is forced to be zero on-shell \cite{Capri:2018un}, implying $A^h_\mu = A_\mu$ and $\psi^h (\bar \psi^h) = \psi (\bar \psi)$. This can also be observed by looking at the form of the Stueckelberg propagator (see below, $\langle \xi(p) \xi(-p)  \rangle= \frac{\alpha}{p^4}$), which vanishes in the Landau gauge.

The lhs of \eqref{gaugeind}, on the other hand, can be expanded using \eqref{Ahexp} for $A^h$ and the corresponding series for $\psi^h$ and $\bar \psi^h$ in terms of the Stueckelberg field. Therefore, one can formally write the identity
\begin{equation} \label{lkft}
\begin{split}
\langle  A_{\mu_1}(x_1) \dots A_{\mu_n}(x_n) \bar \psi (y_1) \psi(z_1) \dots \bar\psi (y_m) \psi(z_m)  \rangle_{\alpha} & = \\
\langle  A_{\mu_1}(x_1) \dots A_{\mu_n}(x_n) \bar \psi (y_1) \psi(z_1) \dots \bar\psi (y_m) \psi(z_m)  \rangle_{\alpha=0} & -  {\cal R}_{\alpha}(x_1,\dots,x_n, y_1, \dots, y_m, z_1, \dots, z_m),
\end{split}
\end{equation}
where ${\cal R}_{\alpha}(x_1,\dots,x_n, y_1, \dots, y_m, z_1, \dots, z_m)$ stands for the remaining expression coming from the expansion of the gauge invariant fields. It represents an infinite series of correlation functions of composite operators, which in principle can be evaluated, at least in perturbation theory at any fixed order, using the propagators and vertices derived from action \eqref{action} with an arbitrary value $\alpha$ for the gauge-fixing parameter.
\subsection{Feynman rules}
In order to carry out an actual evaluation of a relatively simple LKFT (see section below), we need to derive the Feynman rules for the propagators and vertices. The expressions for the tree level propagators, which are the elements of the inverse of the quadratic part of the classical action, were given in \cite{Capri:2018ir}. The non-vanishing ones are given by
\begin{equation} \label{prop}
\begin{split}
\langle A^a_\mu (p) A^b_\nu(-p) \rangle_0 & = \frac{1}{p^2} \left(\delta_{\mu \nu} +(\alpha-1)\frac{p_\mu p_\nu}{p^2} \right) \delta^{ab}, \\
\langle A^a_\mu (p) b^b(-p) \rangle_0 & = \frac{p_\mu}{p^2}\,  \delta^{ab},\\
\langle A^a_\mu (p) \xi^b(-p) \rangle_0 & =-i\alpha \frac{p_\mu}{p^4}\, \delta^{ab}, \\
\langle b^a (p) \xi^b(-p) \rangle_0 & = \frac{i}{p^2} \, \delta^{ab}, \\
\langle \xi^a(p) \xi^b(-p) \rangle_0 & = \frac{\alpha}{p^4} \, \delta^{ab}, \\
\langle \xi^a(p) \tau^b(-p) \rangle_0 & = \langle \bar c^a(p) c^b(-p) \rangle_0 = \langle \bar \eta^a(p) \eta^b(-p) \rangle_0 =\frac{1}{p^2}\, \delta^{ab},
\end{split}
\end{equation}
where the mixing fields propagators result from the fact that the quadratic part in the action is not diagonal. The auxiliary field $b^a$, despite appearing in mixing propagators, does not show up in loop diagrams, since its own propagator vanishes in any gauge and it does not interact with any other field.

Note that in \cite{Capri:2018ir}, besides including the massive gluon parameter which enters the transverse part of the gluon propagator and yields a non-vanishing expression for the  $\tau$ field propagator, an additional dimensional gauge-fixing parameter $\mu$ is introduced in a BRST-exact fashion. It is conveniently used in order to regularize the potentially dangerous infrared divergences originated, in four dimensions, by the dipole ghost like propagator of the dimensionless Stueckelberg field. This extra unphysical scale (see \cite{Capri:2016grib} for a different infrared regularization) enters in the propagators given in \eqref{prop} and adds a non-vanishing expression for $\langle A^a_\mu(p) \tau^b(-p)\rangle$ to that list .

For the evaluation of the one loop LKFT for the gluon propagator, we set this infrared regularizing parameter to zero in order to simplify the calculations. Additionally, the absence of massive parameters allows us to evaluate the one loop integrals with ease in arbitrary spacetime dimensions. The employment of dimensional regularization indeed generates some spurious infrared divergences in the intermediate steps but they eventually cancel out in the final result. We stress that the same calculation has been performed by also including the regularizing parameter $\mu$, and the same result, in $D=4$ dimensions, has been obtained once $\mu$ is sent to zero in the final expression. Said otherwise, there is no need for an infrared regularization/renormalization in our calculation.

In order to obtain the LKFT for the gluon propagator at the one loop level, we need to keep terms up to ${\cal O}(g^2)$
in the expansion of $A^h_\mu$ in \eqref{Ahexp} and derive  Feynman rules for the those tree level vertices which add to the usual vertices of QCD arising from the action $S_h$ in \eqref{action}. These have been obtained using the {\sf Mathematica} package {\sf FeynRules} \cite{Alloul:2014}, and are given by (a total momentum conservation is implicitly understood)
\begin{equation}
\begin{split}
\langle A^a_\mu(p) \tau^b(q) \xi^c(k) \rangle_0 & = -ig f^{abc} q_\mu, \\
\langle \tau^a(p) \xi^b(q) \xi^c(k) \rangle_0 & = \frac{g}{2} f^{abc}\, p{ \bf{\cdot}} (k-q), \\
\langle \bar \eta^a(p) \eta^b(q) A^c_\mu(k) \rangle_0 & = i g f^{abc} p_\mu, \\
\langle \bar \eta^a(p) \eta^b(q) \xi^c(k) \rangle_0 & = -g f^{abc} p{ \bf{\cdot}} k, \\
\langle A^a_\mu(p) \tau^b (q) \xi^c(k) \xi^d (l) \rangle_0 & = i \frac{g^2}{2} q_\mu \left(f^{ade}f^{ebc} + f^{ace}f^{ebd} \right), \\
\langle \tau^a(p) \xi^b(q) \xi^c(k) \xi^d(l) \rangle_0 & = \frac{g^2}{6} \left[f^{ade}f^{ebc} p{\bf \cdot}(k-q) + f^{ace}f^{ebd}p{\bf \cdot}(l-q)+f^{abe}f^{ecd}p {\bf \cdot}(l-k) \right].
\end{split}
\end{equation}
\section{LKFT for the gluon propagator} \label{lkftgluon}
The transformation \eqref{lkft} for the case of 2-point gauge fields, where the remaining part is obtained by expanding $A^h_\mu$ up to
${\cal O}(g^2)$, reads
\begin{eqnarray}
\langle A^a_\mu(x) A^b_\nu (y) \rangle_\alpha  &=& \langle A^a_\mu(x) A^b_\nu (y) \rangle_{\alpha=0} +2\langle A^a_\mu(x) \partial_\nu \xi^b (y) \rangle_\alpha - \langle \partial_\mu \xi^a(x) \partial_\nu \xi^b(y) \rangle_\alpha \nonumber \\
&+& 2gf^{bcd}\langle A^a_\mu(x) A^{c}_\nu(y) \xi^d(y) \rangle_\alpha -2gf^{bcd} \langle \partial_\mu \xi^a(x) A^c_\nu(y) \xi^d(y) \rangle_\alpha \nonumber\\
&-& g^2 f^{ace}f^{bdf} \langle A^c_\mu(x) \xi^e(x) A^d_\nu(y) \xi^f(y) \rangle_\alpha + g f^{bcd} \langle A^a_\mu(x) \xi^c(y) D^{de}_\nu \xi^e (y) \rangle_\alpha  \nonumber \\
&-& g f^{bcd} \langle \partial_\mu \xi^a(x) \xi^c(y) D^{de}_\nu \xi^e (y)  \rangle_\alpha -g^2 f^{ace}f^{bdf} \langle A^c(x)\xi^e(x)\xi^d(y)\partial_\nu \xi^f(y) \rangle_\alpha \nonumber \\
&-& \frac{g^2}{4}f^{ace}f^{bdf} \langle \xi^c(x) \partial_\mu \xi^e(x) \xi^d(y) \partial_\nu \xi^f(y) \rangle_\alpha + \frac{g^2}{3}f^{bce}f^{edf}\langle A^a_\mu(x) \xi^c(y) \xi^d(y) \partial_\nu \xi^f(y) \rangle_\alpha  \nonumber \\
&-& \frac{g^2}{3}f^{bce}f^{edf}\langle \partial_\mu \xi^a(x) \xi^c(y) \xi^d(y) \partial_\nu \xi^f(y) \rangle_\alpha + {\cal O}(g^3),
\label{lkftgl}
\end{eqnarray}
where the symmetry under the interchange $a\leftrightarrow b$, $\mu \leftrightarrow \nu$, $x \leftrightarrow y$ has been employed (e.g. $\langle A^a_\mu(x) \partial_\nu \xi^b(y) \rangle = \langle \partial_\mu \xi^a(x) A^b_\nu(y) \rangle$). The last two terms in the first line of \eqref{lkftgl}, evaluated at zero order, yield the longitudinal part of the gluon propagator at tree level. In fact, Fourier transforming to momentum space ($\partial_\mu \rightarrow -i p_\mu$) and substituting the Feynman rules given in \eqref{prop} for the $A_\mu \text{-} \xi$ and $\xi$ correlation functions, we readily see that these two terms add up to $\alpha\, p_\mu p_\nu/p^4$. Note that in the Abelian case this represents the whole content of the LKFT, since all the other terms in \eqref{lkftgl} vanish and there are no quantum corrections beyond tree level for the terms in the first line. It is due to the fact that there are no interactions among the extra fields, as it is easily seen setting the structure constants to zero. Though the LKFT for the photon propagator is rather trivial, it encodes the fundamental property of gauge independence of the vacuum polarization, which is related to the physical character of the effective electric charge, and the non-dressing of the unphysical longitudinal part of the propagator. This conclusion can also be inferred from the Ward identity. The longitudinal part remains
undressed in the non-Abelian case. However, the transverse part acquires a non-trivial dependence on the gauge parameter which we obtain by the explicit evaluation of the correlation functions in \eqref{lkftgl} to ${\cal O}(g^2)$.

The corresponding Feynman diagrams have been generated by implementing the model with {\sf FeynArts} \cite{Hahn:2000} \footnote{We are specially grateful to Thomas Hahn for fixing a bug concerning the generation of diagrams involving mixing scalar propagators.}, and evaluated by using the reduction routine of {\sf FeynCalc} \cite{Shtabovenko:2016}. The automatization of the calculation is in order, due to the proliferation of diagrams expected from the presence of the mixing propagators and extra vertices.

The correlation functions below the first line in \eqref{lkftgl} are expectation values of composite operators. They include at least two fields evaluated at the same point. The practical way to calculate the corresponding Feynman diagrams is to introduce external sources attached to these composite operators and then evaluate the 2-point correlation functions involving one (or more) extra source which takes care of inserting the corresponding composite operator. Therefore, to the classical action, we add a term that includes three external sources attached to the three composite operators which appear in the expansion of $A^h_\mu$ \eqref{Ahexp} up to ${\cal O}(g^2)$ considered here
\begin{equation}
S_{J} = -\int d^D x \left [ gf^{abc}\xi^b A_\mu^c J_\mu^{1, a} -\frac{g}{2}f^{abc}\xi^b D^{cd}_\mu \xi^d J^{2,a}_\mu  -\frac{g^2}{3!} f^{abd}f^{dce} \xi^b \xi^c \partial_\mu \xi^e J^{3,a}_\mu \right].
\end{equation}
It yields additional vertices with the following Feynman rules
\begin{equation}
\begin{split}
\langle J^{1,a}_\mu(p) A^b_\nu(q) \xi^c(k) \rangle_0 & = -g f^{abc} \delta_{\mu \nu}, \\
\langle J^{2,a}_\mu(p) \xi^b(q) \xi^c(k) \rangle_0 & = i\frac{g}{2} f^{abc} (k_\mu-q_\mu), \\
\langle J^{2,a}_\mu(p)  A^b_\nu(q) \xi^c(k) \xi^d(l) \rangle_0 & = -\frac{g^2}{2}\left(f^{ace}f^{bde}+f^{ade}f^{bce} \right) \delta_{\mu \nu}, \\
\langle J^{3,a}_\mu(p) \xi^b(q) \xi^c(k) \xi^d(l)\rangle_0 & = i\frac{g^2}{6}\left[f^{ade}f^{bce}(k_\mu-q_\mu)+f^{ace}f^{bde}(l_\mu-q_\mu)+f^{abe}f^{cde}(l_\mu-k_\mu) \right].
\end{split}
\end{equation}
In order to illustrate the details of the computation, we focus exclusively on the non-trivial transverse part of the LKFT. We leave the discussion of the longer longitudinal part to the Appendix, where it is found that these contributions sum up to zero beyond tree level, as dictated by the Ward identity. This serves as an important consistency check on both methodology and explicit computation.

The transverse part involves considerably fewer diagrams than the longitudinal one. It owes itself to the fact that the external mixing scalar-vector propagators, being proportional to the external momentum, do not show up. Moreover, the correlation functions in \eqref{lkftgl} involving a derivative of an elementary field, like the ones in the first line, also contribute only to the longitudinal part once Fourier transformed. However, note that if a derivative is part of a composite operator, that correlation function can yield transverse contributions, since, once Fourier transformed, the derivative translates to a loop momentum instead of an external one.

We also stress that at the one loop level the new ghost fields do not appear in the transverse part of the LKFT, while they play an important role in cancelling longitudinal contributions (see Appendix).

\begin{figure}
\hspace*{-5cm}
\begin{subfigure}{.8\textwidth}
\centering
\includegraphics[page=1, width=0.9 \paperwidth]{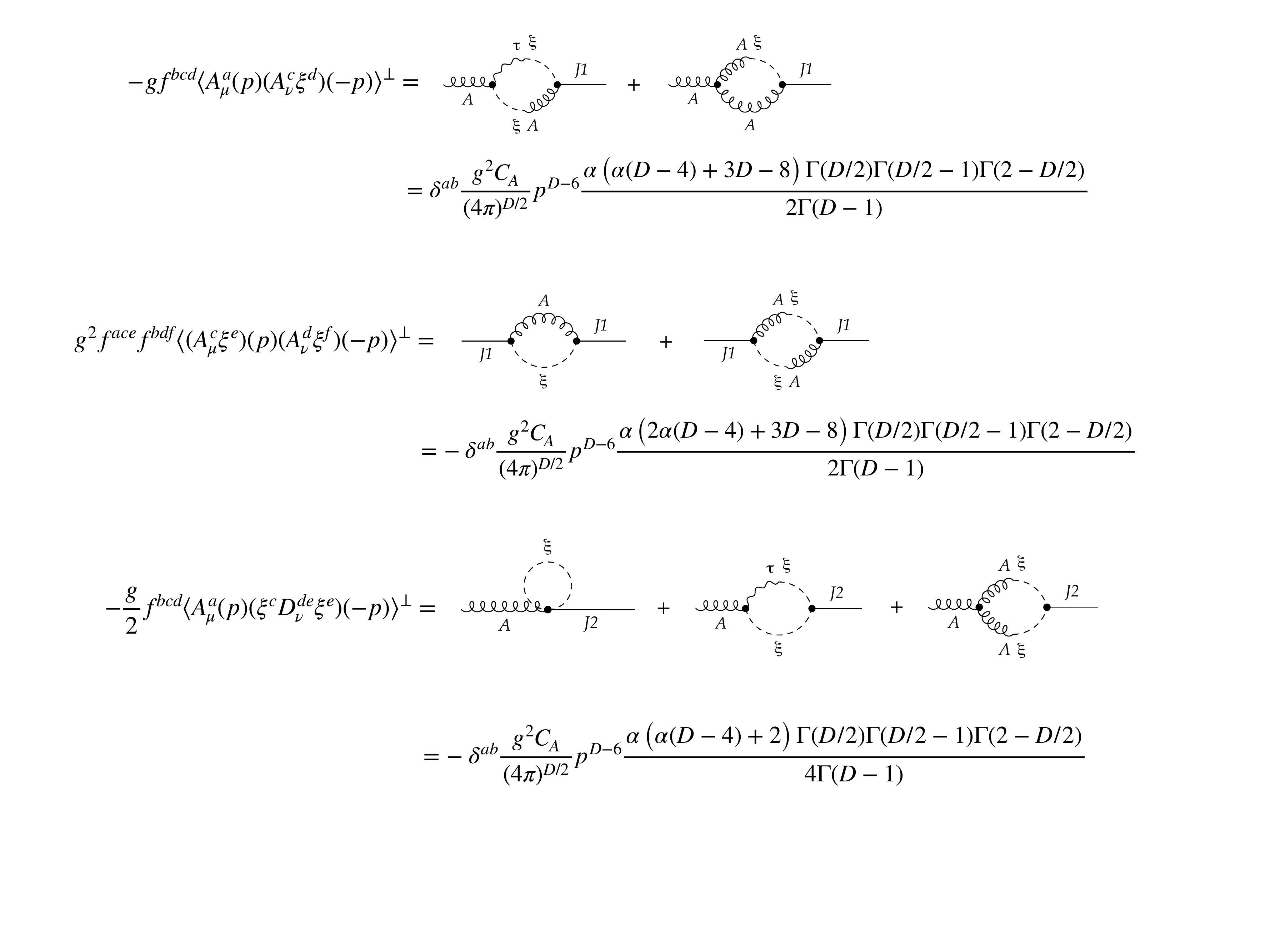}
\end{subfigure}
\hspace*{-3cm}
\begin{subfigure}{.8\textwidth}
\vspace{-2.5cm}
\includegraphics[page=2, width=0.9 \paperwidth]{lkfdiag}
\end{subfigure}
\newline
\vspace{-4.5cm}
\caption{Transverse parts of different correlation functions that enter the LKFT of the gluon propagator (see eq. \eqref{lkftgl}) at the one loop level in $D$ spacetime dimensions.}
\label{lkftrans}
\end{figure}
In Fig.\ref{lkftrans}, we show the Feynman diagrams and their corresponding expressions in $D$ spacetime Euclidean dimensions which contribute to the transverse part of the correlation functions in \eqref{lkftgl} at the one loop level. The upper symbol $\perp$ on the correlation functions stands for the transverse part, meaning that their expressions have been contracted with $(\delta_{\mu \nu}-p_\mu p_\nu/p^2)/(D-1)$. Putting all the pieces together, the final result for the LKFT of the gluon propagator in momentum space is\footnote{Here we have used the result, proved in the Appendix, that all the longitudinal contributions cancel out beyond the tree level.}
\begin{equation}
\begin{split}
\langle A^a_\mu(p) A^b_\nu (-p) \rangle_\alpha  = &\langle A^a_\mu(p) A^b_\nu (-p) \rangle_{\alpha=0}+\alpha \frac{p_\mu p_\nu}{p^4} \delta^{ab} \\
& -\left(\delta_{\mu \nu}-\frac{p_\mu p_\nu}{p^2} \right)\delta^{ab} \frac{g^2 C_A}{(4\pi)^{D/2}}p^{D-6}\frac{\alpha(\alpha(D-4)+6D-20)\Gamma^2(D/2)\Gamma(2-D/2)}{2(D-2)\Gamma(D-1)}.
\end{split}
\end{equation}
It coincides with the known result obtained from a direct evaluation in a generic linear covariant gauge (see for instance \cite{Grozin:2005}) in
the usual formalism of QCD.

\section{Nielsen identities}

We recall here that the Nielsen identities \cite{Nielsen:1975fs} can also be used to derive an explicit relationship of the variation of any Green function under a gauge parameter change, see also \cite{Gambino:1999ai,Capri:2016gut}. Therefore, we expect the LKFTs and the Nielsen identities to have common grounds and origins. We uncloak this overlap by employing the Slavnov-Taylor identity, which itself expresses the BRST invariance of the underlying theory. For the record, in \cite{Quadri:2014jha} the problem of the gauge dependence of 1-PI correlators has been approached via a canonical transformation explicitly solving the extended
Slavnov-Taylor identity.

\subsection{BRST invariance and its extended version}
The action given in \eqref{action} enjoys exact BRST nilpotent symmetry, $sS=0$, with
\begin{eqnarray}
sA^{a}_{\mu}&=&-D^{ab}_{\mu}c^{b}\,,\nonumber \\
s\psi^{i}&=&-igc^{a}(T^{a})^{ij}\psi^{j}\,,\nonumber \\
s{\bar\psi}^{i}&=&igc^{a}\bar{\psi}^{j}(T^{a})^{ji}\,, \nonumber \\
sc^{a}&=&\frac{g}{2}f^{abc}c^{b}c^{c}\,, \nonumber \\
s\bar{c}^{a}&=&ib^{a}\,,\nonumber \\
sb^{a}&=&0\,, \nonumber \\
s \tau^a & = & 0\,, \nonumber \\
s {\bar \eta}^a & = & s \eta^a = 0 \,.
\label{brst}
\end{eqnarray}
The BRST operator $s$ is nilpotent, {\it i.e.}
\begin{equation}
s^2=0 \;. \label{np}
\end{equation}
\subsection{Extended BRST invariance}
One notices that the Faddeev-Popov gauge fixing term, eq.\eqref{eq:gf}, can be rewritten as an exact BRST variation, namely
\begin{equation}
S_{FP} =  \int d^4x \;s \left( {\bar c}^a (\partial_\mu A^a_\mu  - {i} \frac{\alpha}{2} {\bar c^a} b^a \right)     \;. \label{gf1}
\end{equation}
This important property has allowed the authors \cite{Piguet:1984js} to extend the variation of the operator $s$ on the gauge parameter $\alpha$ itself, by keeping nilpotency, exact BRST invariance and renormalizability, namely
\begin{equation}
s \alpha = \chi \;, \qquad s\chi = 0 \;,   \qquad s^2=0 \;,   \label{exts}
\end{equation}
where $\chi$ is a Grassmann parameter with ghost number 1, which is to be set to zero at the end. Expression \eqref{gf1} now reads
\begin{equation}
S_{FP} =  \int d^4x \; \left( ib^{a}  \,\partial_{\mu}A^{a}_{\mu}+ \frac{\alpha}{2} b^a b^a
+\bar{c}^{a}\,\partial_{\mu}D^{ab}_{\mu}c^{b}  - {i} \frac{\chi}{2} {\bar c^a} b^a \right)  \;.
\end{equation}
Evidently, due to the nilpotent character of the extended BRST operator, the action $S$ remains BRST invariant. Nevertheless, as we shall see
shortly, the use of the extended BRST transformation, eq.\eqref{exts}, will allow us to derive both LKF transformations and Nielsen identities from the generating functional of the theory in a very elegant and powerful way.

\subsection{A generalized Slavnov-Taylor identity}

We are now ready to translate the exact BRST invariance into the functional form of the Slavnov-Taylor identity, which will enable us to establish useful properties of  the Green functions of the  theory.  Following the algebraic setup of \cite{Piguet:1995er},  we need to introduce a set of external BRST invariant sources $(\Omega^a_\mu, \bar{\mathcal{L}}^{i}, \mathcal{L}^{i}, L^a, K^a)$ coupled to the non-linear BRST variations of the fields $(A^a_\mu, \psi^{i}, \bar{\psi}^{i}, c^a, \xi^{a})$ as well as the sources $(\mathcal{J}_{\mu}^{a}, \bar{\mathcal{T}^{i}}, \mathcal{T}^{i})$ coupled to the BRST invariant  composite operators $(A_{\mu}^{ha}, \psi^{hi}, \bar{\psi}^{hi})$, eq.\eqref{Ahloc},
\begin{equation}
s \Omega^a_\mu = s \mathcal{L}^{i} = s \bar{\mathcal{L}^{i}} = s L^a = sK^a = s \mathcal{J}_{\mu}^{a}  = s\bar{\mathcal{T}^i} = s \mathcal{T}^i = 0 \;.   \label{invs}
\end{equation}
We shall thus start with the BRST invariant complete action $\Sigma$ defined by \footnote{Where the function $g^{ab}\left(\xi\right)$ should not be confused with the coupling strength $g$. The expression for this power series in $\xi$ can be found in, for instance, \cite{Capri:2016a2}}
\begin{eqnarray}
\Sigma & = & \int d^{4}x\left( \frac{1}{4}\left(F_{\mu\nu}^{a}\right)^{2} + \bar{\psi}(\slashed{D}+m_f)\psi +ib^{a}\partial_{\mu}A_{\mu}^{a}+\bar{c}^{a}\partial_{\mu}D_{\mu}^{ab}c^{b}+\frac{\alpha}{2}\left(b^{a}\right)^{2} - {i} \frac{\chi}{2} {\bar c^a} b^a   \right. \nonumber \\
&  & +\tau^{a}\partial_{\mu}A_{\mu}^{ha} +\bar{\eta}^{a}\partial_{\mu}D_{\mu}^{ab}\left(A^{h}\right)\eta^{b}+\frac{m^2}{2}A_{\mu}^{ha}A_{\mu}^{ha} + \mathcal{J}_{\mu}^{a}A_{\mu}^{ha} + \bar{\mathcal{T}^{i}} \psi^{hi} +\mathcal{T}^i\bar{\psi}^{hi}\nonumber \\
&  & -\Omega_{\mu}^{a}D_{\mu}^{ab}c^{b}-ig\bar{\mathcal{L}^{i}}c^{a}(T^{a})^{ij}\psi^{j}+ig\mathcal{L}^{i}c^a(T^{a})^{ji}\bar{\psi}^{j}+\frac{g}{2}f^{abc}L^{a}c^{b}c^{c}+K^{a}g^{ab}\left(\xi\right)c^{b} \Bigl)  \;.  \label{cact}
\end{eqnarray}
The BRST invariance of the external sources ensures that
\begin{equation}
s \Sigma = 0 \;. \label{Sg}
\end{equation}
Therefore, the complete action $\Sigma$ obeys the following Slavnov-Taylor identity
\begin{equation}
\mathcal{S}\left(\Sigma\right)  =  \int d^{4}x\left(\frac{\delta \Sigma}{\delta\Omega_{\mu}^{a}}\frac{\delta \Sigma}{\delta A_{\mu}^{a}}+\frac{\delta \Sigma}{\delta \bar{\mathcal{L}}^{i}}\frac{\delta \Sigma}{\delta \psi^{i}}+\frac{\delta \Sigma}{\delta \mathcal{L}^{i}}\frac{\delta \Sigma}{\delta \bar{\psi}^{i}}+\frac{\delta \Sigma}{\delta L^{a}}\frac{\delta \Sigma}{\delta c^{a}}+ib^{a}\frac{\delta \Sigma}{\delta\bar{c}^{a}}+\frac{\delta \Sigma}{\delta K^{a}}\frac{\delta \Sigma}{\delta\xi^{a}}\right)  + \chi \frac{\delta \Sigma}{\delta \alpha}= 0  \;.   \label{sti}
\end{equation}
The all order renormalizability of the action \eqref{cact} can be established within the algebraic renormalization framework without any difficulty by repeating the analysis already performed in \cite{Fiorentini:2016rwx}. We shall thus directly focus on the consequences which can be derived from the generalized Slavnov-Taylor identity \eqref{sti}. We rewrite it for the generator $\Gamma$ of the 1PI Green functions of the theory:
\begin{equation}
\mathcal{S}\left(\Gamma\right)  =  \int d^{4}x\left(\frac{\delta \Gamma}{\delta\Omega_{\mu}^{a}}\frac{\delta \Gamma}{\delta A_{\mu}^{a}}+\frac{\delta \Gamma}{\delta \bar{\mathcal{L}}^{i}}\frac{\delta \Gamma}{\delta \psi^{i}}+\frac{\delta \Gamma}{\delta \mathcal{L}^{i}}\frac{\delta \Gamma}{\delta \bar{\psi}^{i}}+\frac{\delta \Gamma}{\delta L^{a}}\frac{\delta \Gamma}{\delta c^{a}}+ib^{a}\frac{\delta \Gamma}{\delta\bar{c}^{a}}+\frac{\delta \Gamma}{\delta K^{a}}\frac{\delta \Gamma}{\delta\xi^{a}}\right)  + \chi \frac{\delta \Gamma}{\delta \alpha}= 0  \;.  \label{stiq}
\end{equation}
For later convenience, it is helpful to already perform the Legendre transformation and rewrite the Slavnov-Taylor identity  \eqref{stiq} for the generator of the connected Green function $Z[J,\mathcal{Q},\mu]$
\begin{equation}
Z[J,\mathcal{Q},\mu]=\Gamma[\Phi,\mathcal{Q},\mu]+\sum_{i}\int d^{4}x J^{(\Phi)}_{i}\Phi_{i}\,,    \label{zj}
\end{equation}
whereby $J$ stands for the standard sources coupled to the fields of the theory and
\begin{eqnarray}
\Phi_i&\equiv&\{A_\mu,\psi,\bar{\psi},b,c,\bar{c},\xi, \tau,\eta,\bar\eta \}\,,\nonumber\\
\mathcal{Q}&\equiv&\{\Omega_\mu,\bar{\mathcal{L}}, \mathcal{L}, L,K, {\cal J}_\mu, \bar{\mathcal{T}}, \mathcal{T}\}\,,\nonumber\\
\mu&\equiv&\{\chi,\alpha\}\,.
\end{eqnarray}
From expression \eqref{zj} we have
\begin{equation}
\begin{tabular}{rclrcl}
$\displaystyle \frac{\delta\Gamma}{\delta\Phi^{bos}_{i}}$&$=$&$-J^{(\Phi^{bos})}_{i}\,,\qquad$&$\displaystyle\frac{\delta Z}{\delta J^{(\Phi^{bos})}_{i}}$&$=$&$\Phi^{bos}_{i}\,,$
\end{tabular}
\end{equation}
for bosonic fields and
\begin{equation}
\begin{tabular}{rclrcl}
$\displaystyle \frac{\delta\Gamma}{\delta\Phi^{fer}_{i}}$&$=$&$J^{(\Phi^{fer})}_{i}\,,\qquad$&$\displaystyle\frac{\delta Z}{\delta J^{(\Phi^{fer})}_{i}}$&$=$&$\Phi^{fer}_{i}\,,$
\end{tabular}
\end{equation}
for the fermionic ones. Also
\begin{equation}
\begin{tabular}{rclrcl}
$\displaystyle \frac{\delta\Gamma}{\delta\mathcal{Q}}$&$=$&$\displaystyle \frac{\delta Z}{\delta\mathcal{Q}}\,,\qquad$
&$\displaystyle \frac{\partial\Gamma}{\partial\mu}$&$=$&$\displaystyle \frac{\partial Z}{\partial\mu}\,.$
\end{tabular}
\end{equation}
When written in terms of the generating functional $Z[J,\mathcal{Q},\mu]$, the previous identity, eq.\eqref{stiq},  takes the form:
\begin{equation}  \label{ST_Z}
 \begin{split}
\int d^{4}z \Bigg[ &
-J^{(A^a)}_{\mu}(z)\frac{\delta Z}{\delta\Omega^{a}_{\mu}(z)}
-J^{(\psi^{i})}(z) \frac{\delta Z}{\delta \bar{\mathcal{L}}^{i}(z)}
-J^{(\bar{\psi}^{i})}(z) \frac{\delta Z}{\delta \mathcal{L}^{i}(z)}
+J^{(c^a)}(z)\frac{\delta Z}{\delta L^{a}(z)}  \\
&+iJ^{(\bar{c}^a)}(z)\frac{\delta Z}{\delta J^{(b^a)}(z)}
-J^{(\xi^a)}(z)\frac{\delta Z}{\delta K^{a}(z)}
  \Bigg]
+\chi\,\frac{\partial Z}{\partial \alpha}
=0 \;.
\end{split}
\end{equation}
We are now ready to exploit equation \eqref{ST_Z} at the level of the Green functions of the theory.

As we shall see, the identity \eqref{ST_Z} succinctly encodes both LKF transformations and Nielsen  identities.

\subsection{Derivation of the LKF transformations}
In order to derive the LKF transformations for the connected correlation functions of $n$ gluon fields and $m$ pairs of fermion and anti-fermion fields, we act on the identity \eqref{ST_Z} with the test operator
\begin{equation}
\frac{\partial}{\partial \chi}  \frac{\delta}{\delta{\cal J}_{\mu_1}^{a_1}(x_1)} \cdots   \frac{\delta}{{\delta \cal J}_{\mu_n}^{a_n}(x_n)} \frac{\delta^2}{ \delta {\cal T}^{i_1}(y_1) \delta {\cal \bar  T}^{j_1}(z_1)} \cdots   \frac{\delta^2}{\delta {\cal T}^{i_m }(y_m) \delta {\cal \bar  T}^{j_m}(z_m) }\label{test1}   \;,
\end{equation}
and set all sources $(J,\mathcal{Q})$ and the parameter $\chi$ equal to zero at the very end. We thus arrive at the result
\begin{equation}
\frac{\partial}{\partial \alpha} \langle  A_{\mu_1}^{h{a_1}}(x_1) \dots A_{\mu_n}^{h{a_n}}(x_n) \bar \psi^{h i_1} (y_1) \psi^{h j_1}(z_1) \dots \bar\psi^{h i_m} (y_m) \psi^{h j_m}(z_m)  \rangle = 0 \;, \label{lkf1}
\end{equation}
expressing the gauge parameter $\alpha$ independence of the BRST invariant Green function
\begin{eqnarray*}
 \langle  A_{\mu_1}^{h{a_1}}(x_1) \dots A_{\mu_n}^{h{a_n}}(x_n) \bar \psi^{h i_1} (y_1) \psi^{h j_1}(z_1) \dots \bar\psi^{h i_m} (y_m) \psi^{h j_m}(z_m)  \rangle.
\end{eqnarray*}
From eq.\eqref{lkf1}, we immediately deduce
\begin{equation}
\begin{split}
& \langle  A_{\mu_1}^{h{a_1}}(x_1) \dots A_{\mu_n}^{h{a_n}}(x_n) \bar \psi^{h i_1} (y_1) \psi^{h j_1}(z_1) \dots \bar\psi^{h i_m} (y_m) \psi^{h j_m}(z_m)  \rangle_{\alpha}  \\
= \; &\langle  A_{\mu_1}^{h{a_1}}(x_1) \dots A_{\mu_n}^{h{a_n}}(x_n) \bar \psi^{h i_1} (y_1) \psi^{h j_1}(z_1) \dots\bar \psi^{h i_m} (y_m) \psi^{h j_m}(z_m)  \rangle_{\alpha=0}   \\
= \; &\langle  A_{\mu_1}^{h{a_1}}(x_1) \dots A_{\mu_n}^{h{a_n}}(x_n) \bar \psi^{h i_1} (y_1) \psi^{h j_1}(z_1) \dots \bar\psi^{h i_m} (y_m) \psi^{h j_m}(z_m)  \rangle_{Landau} \;. \label{lkf2}
\end{split}
\end{equation}
Let us also remind ourselves that, in the Landau gauge, the Stueckelberg field $\xi^a$ enjoys the important property of being completely decoupled from the theory, as one can directly realise from its propagator, $\langle \xi(p) \xi(-p)\rangle = \frac{\alpha}{p^4}$, which vanishes when $\alpha=0$. As a consequence, eq.\eqref{lkf2}, takes the simpler form
\begin{equation}
\begin{split}
& \langle  A_{\mu_1}^{h{a_1}}(x_1) \dots A_{\mu_n}^{h{a_n}}(x_n) \bar \psi^{h i_1} (y_1) \psi^{h j_1}(z_1) \dots\bar \psi^{h i_m} (y_m) \psi^{h j_m}(z_m)  \rangle_{\alpha}  \\
=\;  &\langle  A_{\mu_1}^{a_1}(x_1) \dots A_{\mu_n}^{a_n}(x_n) \bar \psi^{ i_1} (y_1) \psi^{ j_1}(z_1) \dots\bar \psi^{ i_m} (y_m) \psi^{ j_m}(z_m)  \rangle_{Landau} \;,  \label{lkf3}
\end{split}
\end{equation}
where $\langle  A_{\mu_1}^{a_1}(x_1) \dots A_{\mu_n}^{a_n}(x_n) \bar \psi^{ i_1} (y_1) \psi^{ j_1}(z_1) \dots\bar \psi^{ i_m} (y_m) \psi^{ j_m}(z_m)  \rangle_{Landau}$ stands for the $(n+2m)$-point connected correlation function in the Landau gauge.  Expanding the composite fields $\left( A_{\mu}^{h{}}, \psi^h, \bar \psi^h \right)$ in powers of $\xi^a$, eq.(\ref{Ahexp},~\ref{psih_expansion}), one arrives at
\begin{equation}
\begin{split}
&\langle  A_{\mu_1}^{h{a_1}}(x_1) \dots A_{\mu_n}^{h{a_n}}(x_n) \bar \psi^{h i_1} (y_1) \psi^{h j_1}(z_1) \dots\bar \psi^{h i_m} (y_m) \psi^{h j_m}(z_m) \rangle_{\alpha} \\
=\; & \langle   A_{\mu_1}^{a_1}(x_1) \dots A_{\mu_n}^{a_n}(x_n) \bar \psi^{ i_1} (y_1) \psi^{ j_1}(z_1) \dots \bar\psi^{ i_m} (y_m) \psi^{ j_m}(z_m) \rangle_{\alpha} + {\cal R}_{\alpha}(x_1,\dots,x_n, y_1, \dots, y_m, z_1, \dots, z_m) \;, \label{lkf4}
\end{split}
\end{equation}
where ${\cal R}_{\alpha}(x_1,\dots,x_n, y_1, \dots, y_m, z_1, \dots, z_m)$ collects all higher order perturbative contribution coming from the interaction vertices of action \eqref{action}, see  \cite{DeMeerleer:2018txc} for the example of the two-point gauge correlation function.

Finally, we have
\begin{equation}
\begin{split}
&\langle  A_{\mu_1}^{a_1}(x_1) \dots A_{\mu_n}^{a_n}(x_n) \bar \psi^{ i_1} (y_1) \psi^{ j_1}(z_1) \dots \bar\psi^{ i_m} (y_m) \psi^{ j_m}(z_m) \rangle_{\alpha} \\
=\; &  \langle  A_{\mu_1}^{a_1}(x_1) \dots A_{\mu_n}^{a_n}(x_n) \bar \psi^{ i_1} (y_1) \psi^{ j_1}(z_1) \dots\bar \psi^{ i_m} (y_m) \psi^{ j_m}(z_m) \rangle_{Landau} - {\cal R}_{\alpha}(x_1,\dots,x_n, y_1, \dots, y_m, z_1, \dots, z_m)\;, \label{lkf5}
\end{split}
\end{equation}
expressing the non-Abelian LKF transformations recently obtained in \cite{DeMeerleer:2018txc} in a compact form.

\subsection{The Nielsen identities}
Let us now proceed by showing that the same identity, eq.\eqref{ST_Z}, also enables us to derive the Nielsen identities of the $(n+2m)$-point correlation function. To that end, we let the following test operator act on equation \eqref{ST_Z}
\begin{equation}
\frac{\partial}{\partial \chi}  \frac{\delta}{\delta J_{\mu_1}^{A^{a_1}}(x_1)} \cdots   \frac{\delta}{ \delta J_{\mu_n}^{A^{a_n}}(x_n)} \frac{\delta^2}{\delta J^{\bar \psi^{i_1}}(y_1) \delta J^{ \psi^{j_1}}(z_1)} \cdots  \frac{\delta^2}{\delta J^{\bar\psi^{i_m}}(y_m) \delta J^{\psi^{j_m}}(z_m)}  \label{n1}   \;,
\end{equation}
and set all sources $(J,\mathcal{Q})$ and the parameter $\chi$ equal to zero. We thus get (note that we omit the spacetime arguments in the functional derivatives of the rhs)\footnote{The notation $\xcancel{J^i}$ means that the term is not included in the product.}
\begin{equation}
\begin{split}
& \frac{\partial}{\partial \alpha} \langle  A_{\mu_1}^{{a_1}}(x_1)\dots A_{\mu_n}^{{a_n}}(x_n) \bar \psi^{i_1}(y_1)\psi^{j_1}(z_1)\dots \bar\psi^{i_m}(y_m)\psi^{j_m}(z_m) \rangle  \\
&= \sum_{k=1}^n  \frac{\partial}{\partial \chi}   \frac{\delta}{\delta J_{\mu_1}^{A^{a_1}}} \cdots \xcancel{ \frac{\delta}{\delta J_{\mu_k}^{A^{a_k}}} } \cdots \frac{\delta}{\delta J_{\mu_{n}}^{A^{a_{n}}}} \frac{\delta^2}{\delta J^{\bar \psi^{i_1}} \delta J^{ \psi^{j_1}}} \cdots  \frac{\delta^2}{\delta J^{\bar\psi^{i_m}} \delta J^{\psi^{j_m}}} \frac{\delta Z}{\delta\Omega^{a_k}_{\mu_k}}    \\
&+ \sum_{k=1}^m  \frac{\partial}{\partial \chi}   \frac{\delta}{\delta J_{\mu_1}^{A^{a_1}}} \cdots  \frac{\delta}{\delta J_{\mu_{n}}^{A^{a_{n}}}} \frac{\delta^2}{\delta J^{\bar \psi^{i_1}}\delta J^{ \psi^{j_1}}} \cdots \xcancel{\frac{\delta^2}{ \delta J^{\bar\psi^{i_k}}\delta J^{\psi^{j_k}}}}  \cdots  \frac{\delta^2}{\delta J^{\bar\psi^{i_m}} \delta J^{\psi^{j_m}}} \frac{\delta^2 Z}{\delta J^{\bar \psi^{i_k}}\delta \bar{\mathcal{L}}^{j_k}} \\
&+ \sum_{k=1}^m \frac{\partial}{\partial \chi}   \frac{\delta}{\delta J_{\mu_1}^{A^{a_1}}} \cdots  \frac{\delta}{\delta J_{\mu_{n}}^{A^{a_{n}}}} \frac{\delta^2}{\delta J^{\bar \psi^{i_1}} \delta J^{ \psi^{j_1}}} \cdots \xcancel{\frac{\delta^2}{ \delta J^{\bar\psi^{i_k}}\delta J^{\psi^{j_k}}}}  \cdots \frac{\delta^2}{\delta J^{\bar\psi^{i_m}}\delta J^{\psi^{j_m}}} \frac{\delta^2 Z}{\delta \mathcal{L}^{i_k} \delta J^{\psi^{j_k}}}  \;, \label{n2}
\end{split}
\end{equation}
namely
\begin{equation}
\begin{split}
& \frac{\partial}{\partial \alpha} \langle  A_{\mu_1}^{{a_1}}(x_1)\dots A_{\mu_n}^{{a_n}}(x_n) \bar \psi^{i_1}(y_1)\psi^{j_1}(z_1)\dots \bar\psi^{i_m}(y_m)\psi^{j_m}(z_m) \rangle  \\
& = \sum_{k=1}^n \langle A_{\mu_1}^{a_1}(x_1) \dots {\left[\frac{i}{2} \int d^4 z\, \bar c^p (z) b^p(z)\right]} D^{{a_k} b}_{\mu_k}c^k(x_k)  \dots A_{\mu_n}^{a_n}(x_n) \bar \psi^{i_1}(y_1)\psi^{j_1}(z_1)\dots \bar\psi^{i_m}(y_m)\psi^{j_m}(z_m)   \rangle \\
& + \sum_{k=1}^m \langle A_{\mu_i}^{a_1}(x_1)\cdots A_{\mu_n}^{a_n}(x_n) \bar \psi^{i_1}(y_1)\psi^{j_1}(z_1)\dots  {\left[ \frac{i}{2}\int d^4z\,{\bar c^p}(z) {b^p}(z) \right]}\\
&\times{\left[-ig \bar\psi^{i_k}(y_k)\left(T^{a} \right)^{j_k l}\psi^l(z_k)c^a(z_k) \right]} \dots \bar\psi^{i_m}(y_m)\psi^{j_m}(z_m)  \rangle \\
&+ \sum_{k=1}^m \langle A_{\mu_i}^{a_1}(x_1)\cdots A_{\mu_n}^{a_n}(x_n) \bar \psi^{i_1}(y_1)\psi^{j_1}(z_1)\dots  {\left[ \frac{i}{2}\int d^4z\,{\bar c^p}(z) {b^p}(z) \right]}\\
&\times{\left[ig \bar c^a(y_k)\bar \psi^{l}(y_k)\left(T^{a} \right)^{l i_k}\psi^{j_k}(z_k) \right]} \dots \bar\psi^{i_m}(y_m)\psi^{j_m}(z_m)  \rangle
\;. \label{n3}
\end{split}
\end{equation}
For  instance, in the case of the two-point gluon correlation function, equation \eqref{n3} yields the known result
\begin{equation}
\frac{\partial}{\partial \alpha} \langle  A_{\mu}^{{a}}(p) A_{\nu}^{{b}}(-p) \rangle =  \delta^{ab} \;\frac{p_\mu p_\nu}{p^4}  \; \; \label{n55}
\end{equation}
at the first order. In the Abelian case, equation \eqref{n55} holds to all orders, due to the free nature of the ghost fields.

\subsection{Summary and conclusion}

Firstly, we explicitly checked the recently derived non-Abelian LKFTs \cite{DeMeerleer:2018txc} at one-loop order to reproduce the correct gauge parameter dependence of the gluon propagator for its transverse as well as longitudinal projections.
We elucidated the role of the extra fields that were introduced to encode these LKFTs in a local, renormalizable fashion and setup a computational framework based on {\sf FeynArts}, {\sf FeynCalc} and {\sf FeynRules}. It can readily be generalized to the fermion sector and higher orders. Our framework provides a mechanism to deal with composite operators entering the game.

Secondly, we derived both the LKFTs and the Nielsen identities from a unique, generalised Slavnov-Taylor identity, eq.\eqref{ST_Z}. This shows the equivalence of both transformations, which enables us to keep control of the gauge parameter dependence of the Green functions of the theory. Evidently, the final result is the same when evaluated with both methods.

Let us end this analysis with a few concluding remarks and potential applications for future:
\begin{itemize}

	\item When both identities are applied to the gluon propagator, the pole mass of the transverse component of the propagator turns out to be independent of the gauge parameter, see also \cite{Gambino:1999ai,Capri:2016gut}.
	\item Inclusion of fermion fields is almost immediate and straightforward. This is due to the possibility of introducing a gauge invariant spinor field $\psi^h$ \cite{DeMeerleer:2018txc}. As in the case of the gluon propagator, both transformations will give rise to the pole mass independence of the fermion propagator. These results will be presented elsewhere.
	\item The differential form of the Nielsen identities, eq.\eqref{n3}, suggests that the dependence on the gauge parameter $\alpha$ could be exponentiated, as we can explicitly verify in the case of the gluon propagator. This means that the whole dependence on the gauge parameter $\alpha$ would be lifted and isolated into an exponential. Schematically, we would have
	\begin{equation}
	\langle  A_{1} \ldots A_{n} \rangle_{\alpha} = e^{{\cal M}(\alpha)} \; \langle  A_{1} \ldots A_{n} \rangle_{\alpha=0} \;, \label{n5}
	\end{equation}
	an equation which would be rather useful for non-perturbative modelling of (higher order) Green functions. A similar equation should also hold for the LKFT. In future work, we will employ our calculational scheme to gain access to the quantity ${\cal M}(\alpha)$, related to a set of 1PI correlation functions with insertions of composite operators, see the rhs of \eqref{n3}.

	\item We expect that the LKFT/Nielsen identity for the fermion propagator will help us understand better the DCSB in the quark sector, maintaining the complete gauge-independence of the chiral condensate (which can be defined via the fermion propagator), as advocated in \cite{Aslam:2016}. QED has this feature as demonstrated in \cite{Bashir:2007}
and explicitly tested for the chiral symmetry breaking solution in QED$_3$ in the same article. Recall
\begin{equation}\label{condensate}
  \braket{\bar \psi \psi}= \int d^Dp \braket{ \bar\psi(p) \psi(-p)}.
\end{equation}
Formally, using Lorentz (or Euclidean) invariance, we may write for the definition of the chiral condensate
\begin{equation}\label{condensate3}
   \braket{ \bar\psi \psi}\equiv \braket{ \bar\psi(x) \psi(x)}= \frac{1}{V}\int d^D x \braket{ \bar\psi(x) \psi(x)}\,,\qquad V=\text{spacetime volume}\,.
\end{equation}
The final expression of \eqref{condensate3} is gauge invariant. Transformation to Fourier space of it does yield the above relation \eqref{condensate}, which thus also ought to be gauge invariant. 
From \eqref{condensate}, we obviously have
\begin{equation}\label{condensate2}
  \frac{\partial}{\partial \alpha}\braket{\bar \psi \psi}= \int d^Dp \frac{\partial}{\partial \alpha} \braket{ \bar\psi(p) \psi(-p)}
\end{equation}
and the latter integrand can be controlled explicitly via the Nielsen identity. As such, we will investigate whether this integral vanishes or not when $\partial_\alpha \braket{ \bar\psi(p) \psi(-p)}$ is evaluated via the Nielsen identity. Let us also draw attention to the fact that $\braket{\bar \psi^h \psi^h}= \braket{\bar \psi \psi}$.

\end{itemize}

\section*{Acknowledgments}
T.~De Meerleer is supported by a KU Leuven FLOF grant. P.~Dall'Olio is supported by a DGAPA-UNAM grant and is grateful for the hospitality at KU Leuven--Kulak where part of this work was initiated. This study was financed in part by the Coordena\c{c}\~{a}o de Aperfei\c{c}oamento de Pessoal de N\'{i}vel Superior - Brasil (Capes) - Finance Code 001. The Conselho Nacional de Desenvolvimento Cient\'{\i}fico e Tecnol\'{o}gico (CNPq-Brazil), the Funda\c{c}\~{a}o de Amparo \`{a} Pesquisa do Estado do Rio de Janeiro (FAPERJ) and the SR2-UERJ are gratefully acknowledged for financial support. S. P. Sorella is a level PQ-1 researcher under the program Produtividade em Pesquisa-CNPq, 300698/2009-7.
This research was also partly supported by Coordinaci\'on de la Investigaci\'on Cient\'ifica
(CIC) of the University of Michoacan and CONACyT,
Mexico, through Grant nos. 4.10 and CB2014-22117, respectively.


\appendix \section{Longitudinal contributions}
Here we present the expressions for the diagrams that contribute to the longitudinal part of the LKFT \eqref{lkftgl} of the gluon propagator and demonstrate the way they cancel among each other in order to guarantee the transversality of the gluon self-energy, as dictated by the corresponding Ward identity. This is a well-known result in QCD that we need to recover in this framework with extra auxiliary fields.

We first consider the one-loop contributions arising from the terms $\langle A^a_\mu (x)\partial_\nu \xi^b(y)\rangle$ and $\langle\partial_\mu\xi^a(x)\partial_\nu\xi^b(y)\rangle$ from the first line in \eqref{lkftgl}, which, transformed into momentum space, become $ip_\nu \langle A^a_\mu(p) \xi^b(-p) \rangle$ and $p_\mu p_\nu \langle \xi^a(p) \xi^b(-p) \rangle$, respectively. They have several contributions, specially due to the diagrams which can be drawn with different combinations of external mixing propagators. Let us consider the one loop corrections to $\langle A^a_\mu(p)\xi^b(-p) \rangle$. The diagrams with external gluon and Stueckelberg propagators sum up to zero
\begin{equation}
\includegraphics[page=1, width=0.7 \paperwidth]{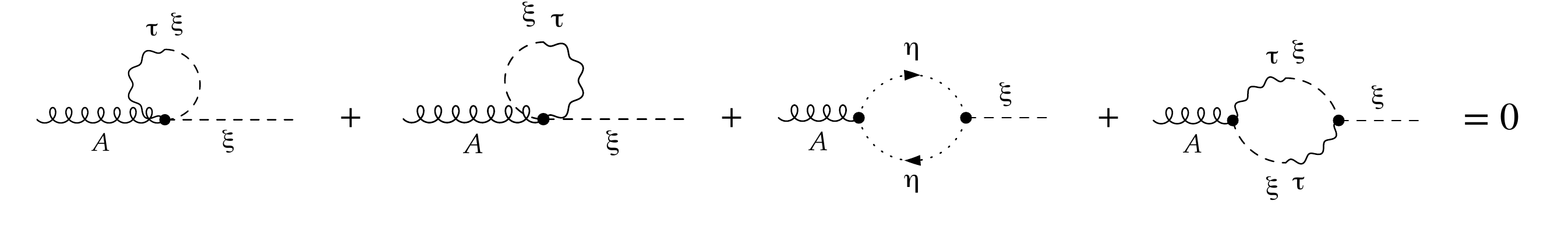}
\end{equation}
The first two diagrams are scaleless tadpoles, which vanish in dimensional regularization, while the last two are non-vanishing and cancel each other out. This cancellation is attributed to the new ghost fields `eating up' diagrams which would otherwise contribute to the longitudinal part of the gauge boson propagator. This result also clearly shows that there are no contributions to the correlation function $\langle \xi^a(p) \xi^b(-p) \rangle$ stemming from diagrams with an external $\xi \text{-}A_\mu$ mixing propagator or a $\xi$ propagator.

We then consider the contributions to $ip_\nu \langle A^a_\mu(p) \xi^b(-p) \rangle$ coming from the diagrams with an external gluon propagator and an external $\tau \text{-} \xi$ propagator. These are given by the following diagrams\footnote{From now on we will omit all the vanishing scaleless tadpole diagrams.}
\begin{equation} \label{ATauXi}
\includegraphics[page=1, width=0.7 \paperwidth]{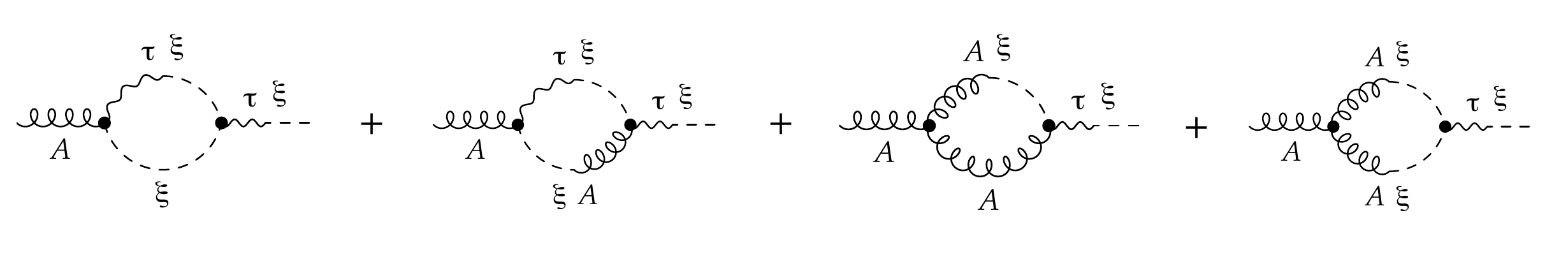}
\end{equation}
Without the need of deriving the explicit expression for these diagrams, it is easy to show that these cancel against the contributions to $p_\mu p_\nu \langle \xi^a(p) \xi^b(-p) \rangle$ coming from the same diagrams but with the external gluon propagator replaced by an external $\xi \text{-} A_\mu$ propagator. In fact, if we denote the sum of the diagrams in \eqref{ATauXi} with the external propagators amputated by $p_\mu \Sigma_{A^a_\mu \tau^b}(p)$, the expression which contributes to $ip_\nu \langle A^a_\mu(p) \xi^b(-p) \rangle$ is
\begin{equation}
ip_\nu \langle A^a_\mu(p) A_\alpha^c(-p) \rangle_0 \left(p_\alpha \Sigma_{A^c_\alpha \tau^d}(p) \right) \langle \tau^d(-p) \xi^b(p) \rangle_0 =i \alpha\frac{p_\mu p_\nu}{p^2}\Sigma_{A^a_\mu \tau^b}(p),
\end{equation}
where we have substituted the expressions for the tree-level propagators given in \eqref{prop}. On the other hand, the expression which contributes to $p_\mu p_\nu  \langle \xi^a(p) \xi^b(-p) \rangle$ is given by
\begin{equation}
p_\mu p_\nu \langle \xi^a(p) A^c_\alpha(-p) \rangle_0 \left(p_\alpha \Sigma_{A^c_\alpha \tau^d}(p) \right)\langle \tau^d(-p) \xi^b(p) \rangle_0 = i \alpha\frac{p_\mu p_\nu}{p^2}\Sigma_{A^a_\mu \tau^b}(p).
\end{equation}
Hence they exactly cancel out inside the gluon LKFT \eqref{lkftgl}. Note that both the contributions appear twice in \eqref{lkftgl}, on one hand because there is a factor of 2 multiplying $ip_\nu \langle A^a_\mu(p) \xi^b(-p) \rangle$, and on the other because there are two topologically inequivalent sets of diagrams which contribute to $p_\mu p_\nu \langle \xi^a(p) \xi^b(-p) \rangle$, one with the mixed $\xi \text{-} A_\mu$ on the left and the mixed $\tau \text{-}\xi$ on the right, while the other with the external propagators swapped.

Exactly the same argument is in place to show the cancellation between the expressions for the diagrams with external $A_\mu \text{-} \xi$ and $\tau \text{-} \xi$ propagators which contribute to $ip_\nu \langle A^a_\mu(p) \xi^b(-p) \rangle$ on one hand, and the diagrams with external $\xi$ and $\tau \text{-} \xi$ propagators which contribute to $p_\mu p_\nu \langle \xi^a(p) \xi^b(-p) \rangle$ on the other. For the former contributions we have
\begin{equation}
i p_\nu \langle A^a_\mu(p) \xi^c(-p) \rangle_0 \left( \Sigma_{\xi^c \tau^d}(p) \right) \langle \tau^d(p) \xi^b(-p) \rangle_0 = \alpha \frac{p_\mu p_\nu}{p^6}\Sigma_{\xi^a \tau^b}(p),
\end{equation}
while for the latter ones
\begin{equation}
p_\mu p_\nu \langle \xi^a(p) \xi^c (-p) \rangle_0  \left( \Sigma_{\xi^c \tau^d}(p) \right) \langle \tau^d(p) \xi^b(-p) \rangle_0 = \alpha \frac{p_\mu p_\nu}{p^6}\Sigma_{\xi^a \tau^b}(p).
\end{equation}
The same exact cancellation does not occur for the contributions coming from the diagrams with two external Stueckelberg propagators in $\langle \xi^a(p) \xi^b(-p) \rangle$ because swapping two identical external propagators does not yield topologically inequivalent diagrams. Fortunately, these diagrams add up to zero:
\begin{equation} \label{XiXi}
\includegraphics[page=1, width=0.38 \paperwidth]{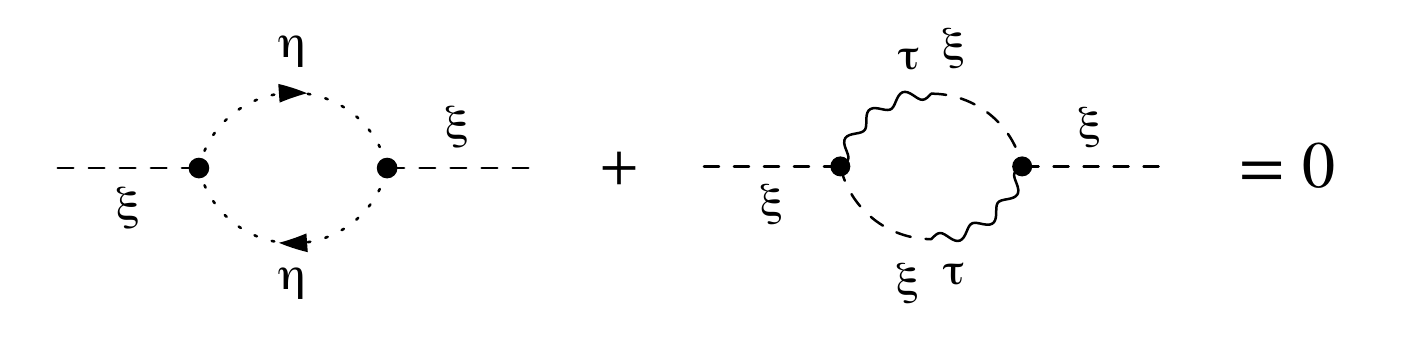}
\end{equation}
Again the new ghost fields serve the purpose of cancelling a loop with internal $\xi \text{-} \tau$ propagators.

There are also contributions coming from diagrams with external gluon and $A_\mu \text{-} \xi$ propagators and with external $\xi \text{-} A_\mu$ and $A_\mu \text{-}\xi$ propagators. The corresponding diagrams with amputated external propagators are the ones which appear in the usual QCD gluon self-energy which yield a transverse expression, plus the following extra diagrams, again involving a loop of ghosts $\eta$ and a loop of mixed $\xi \text{-} \tau$  whose expressions cancel each other out.
\begin{equation} \label{XiAAXi}
\includegraphics[page=1, width=0.41 \paperwidth]{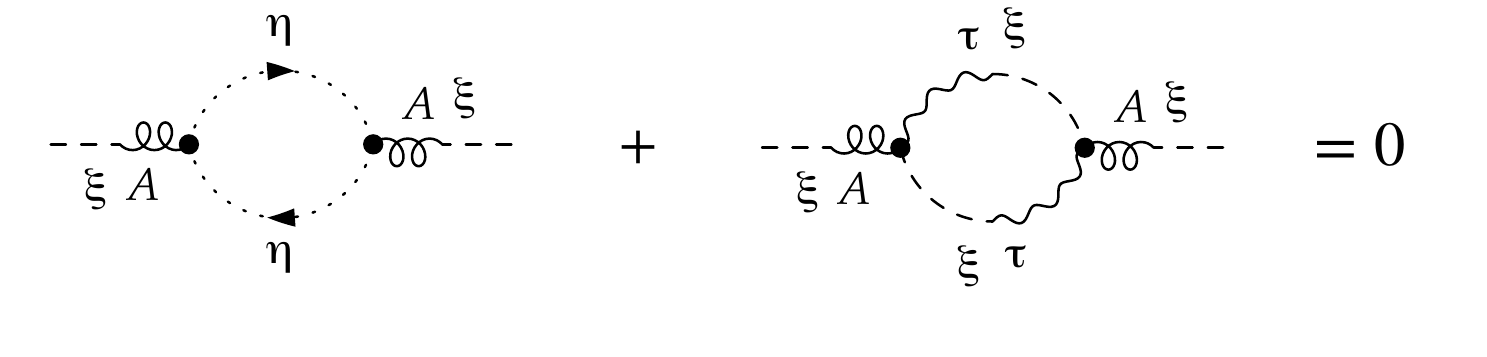}
\end{equation}
There are nevertheless contributions to $p_\mu p_\nu \langle \xi^a(p) \xi^b(-p) \rangle$ which are neither zero nor cancel against corresponding contributions to $ip_\nu \langle A^a_\mu(p) \xi^b(-p) \rangle$. These come from the diagrams with two external $\xi \text{-} \tau$ propagators, which have no counterparts with $A_\mu \text{-} \tau$ propagators, because these propagators are zero in the theory with no explicit infrared regularizing mass parameter. Taking the longitudinal part of these contributions (i.e. $(p_\mu p_\nu \langle \xi^a(p) \xi^b(-p) \rangle)^\parallel = p^2 \langle \xi^a(p) \xi^b(-p) \rangle$, where the upper symbol $\parallel$ stands for the contraction with the longitudinal projector $p_\mu p_\nu /p^2$) one gets
\begin{equation} \label{XiTauTauXi}
\includegraphics[page=1, width=0.75 \paperwidth]{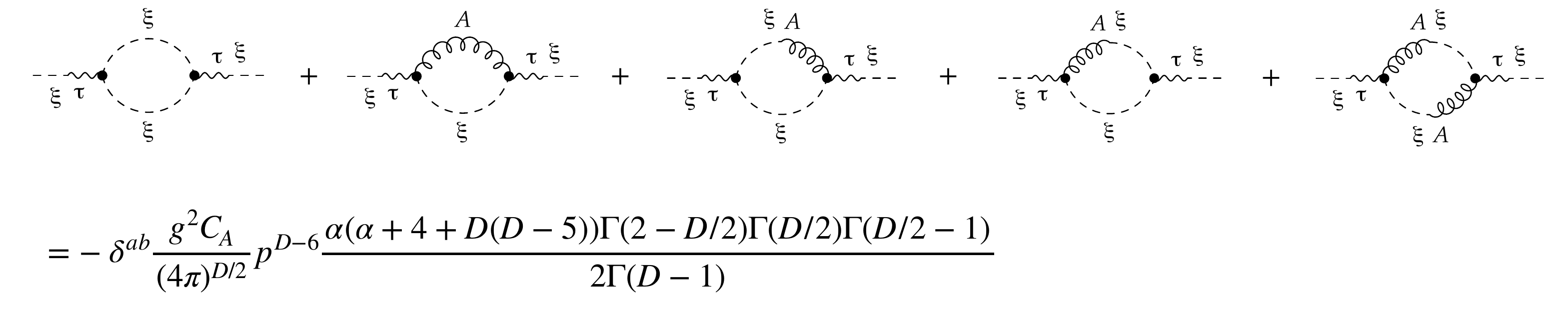}
\end{equation}
This represents the only non-vanishing contribution to the longitudinal part of the LKFT coming from the first line of \eqref{lkftgl}.
It must be cancelled somehow by the correlation functions involving composite operators.
Analogous cancellations occur between diagrams corresponding to $\langle A^a_\mu(p) O^b(-p) \rangle^\parallel$ and $\langle (\partial_\mu \xi^a)(p) O^b(-p) \rangle^\parallel$, where $O(p)$ stands for a composite operator, and the only contributions that survive, for lack of a counterpart, are the diagrams corresponding to $\langle (\partial_\mu \xi^a)(p) O^b(-p) \rangle^\parallel$ with an external $\xi \text{-} \tau$ propagator. These are derived by evaluating the correlation function between the mixed field and the external source attached to the corresponding composite operator (see section \ref{lkftgluon}). For instance, the expression corresponding to $g f^{bcd}\langle (\partial_\mu \xi^a)(p) (A^c_\nu \xi^d)(-p) \rangle^\parallel$ is given by
\begin{equation} \label{XiTauJ1}
\includegraphics[page=1, width=0.55 \paperwidth]{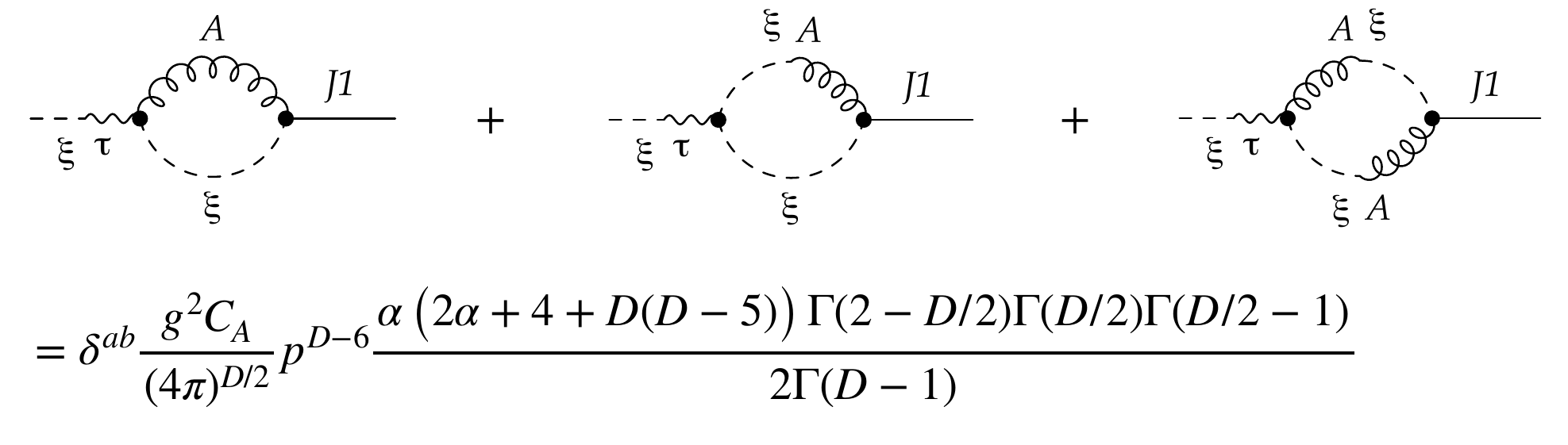}
\end{equation}
and the contribution that survives from $\frac{g}{2} f^{bcd} \langle(\partial_\mu \xi^a)(p) (\xi^c D^{de}_\nu \xi^e)(-p) \rangle^\parallel$ corresponds to
\begin{equation} \label{XiTauJ2}
\includegraphics[page=1, width=0.7 \paperwidth]{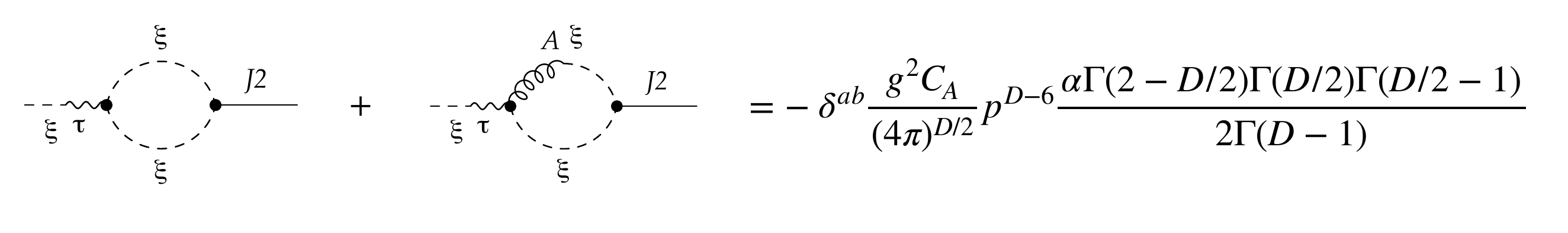}
\end{equation}
What is left to evaluate are the longitudinal contributions to the correlation functions between composite operators. The expression for $g^2 f^{ace}f^{bdf}\langle(A^c_\mu \xi^e)(p)(A^d_\nu \xi^f)(-p) \rangle^\parallel$ is given by
\begin{equation} \label{J1J1}
\includegraphics[page=1, width=0.52\paperwidth]{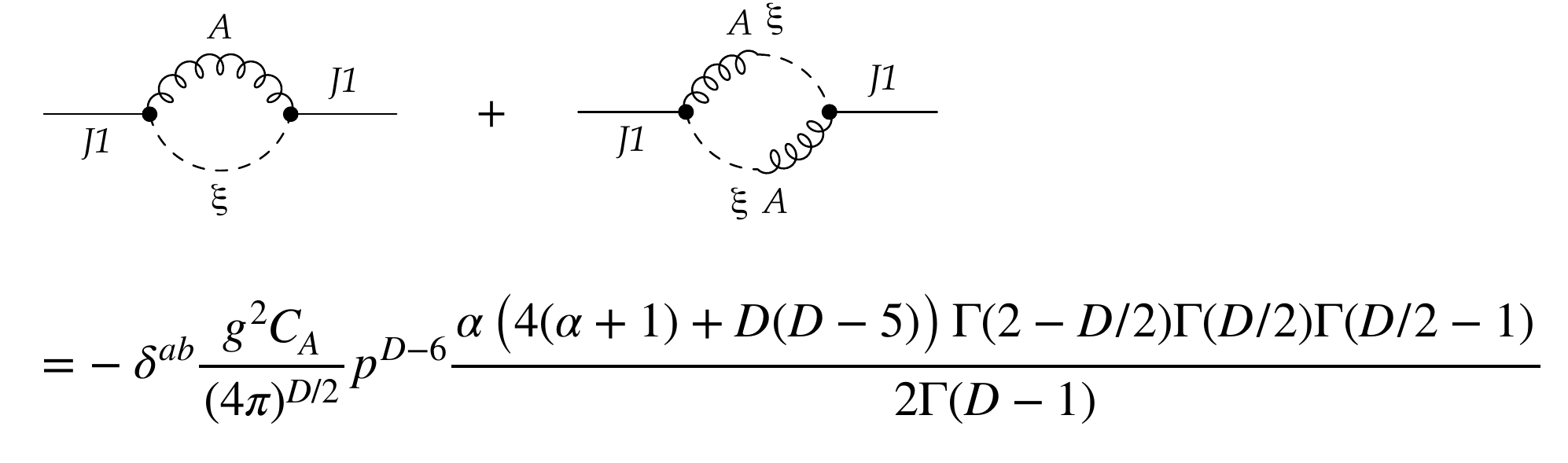}
\end{equation}
The expression for $\frac{g^2}{2}f^{ace}f^{bdf} \langle(A^c \xi^e)(p)(\xi^d \partial_\nu \xi^f)(-p) \rangle^\parallel$ reads as
\begin{equation} \label{J1J2}
\includegraphics[page=1, width=0.52\paperwidth]{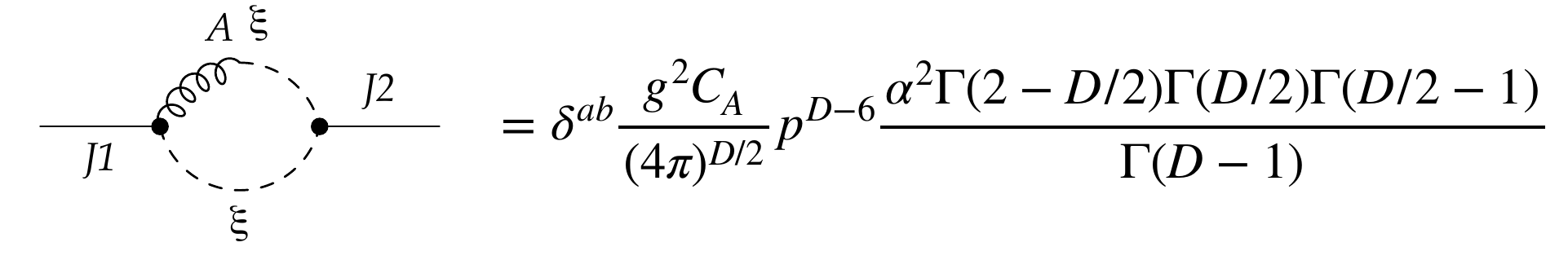}
\end{equation}
Finally, the last contribution to the longitudinal part comes from $\frac{g^2}{4}f^{ace}f^{bdf} \langle (\xi^c \partial_\mu \xi^e)(p) (\xi^d \partial_\nu \xi^f)(-p) \rangle^\parallel$ and is given by
\begin{equation} \label{J2J2}
\includegraphics[page=1, width=0.53\paperwidth]{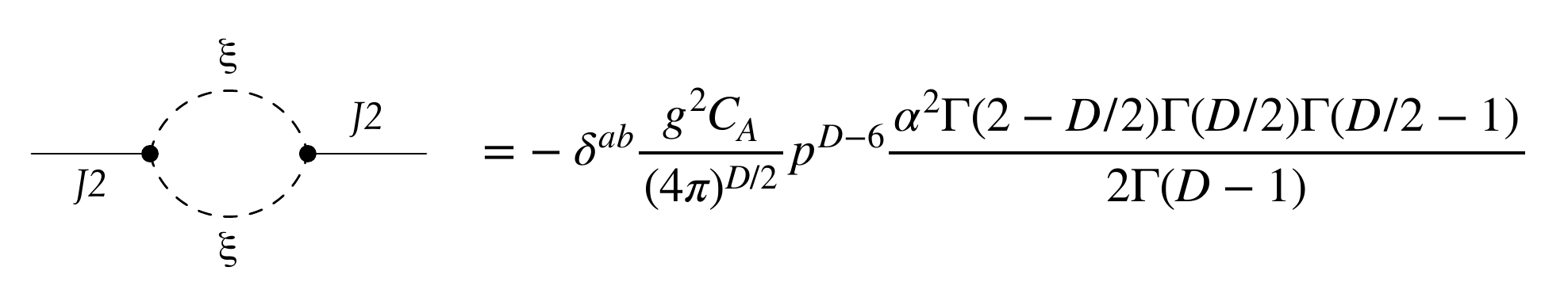}
\end{equation}
Like the transverse part, there are no contributions of ${\cal O}(g^2)$ to $-\frac{g^2}{6}f^{bce}f^{edf} \langle A^a_\mu(p) (\xi^c \xi^d \partial_\nu \xi^f)(-p) \rangle$, nor to $\frac{g^2}{6}f^{bce}f^{edf} \langle \partial_\mu \xi^a(p) (\xi^c \xi^d \partial_\nu \xi^f)(-p) \rangle$.

Adding all the non-zero longitudinal contributions, it is straightforward to see that the final result is zero, as expected (note that the expressions \eqref{XiTauJ1}, \eqref{XiTauJ2} and \eqref{J1J2} have to be multiplied by a factor of 2, because of the duplication of crossing terms inside the LKFT).


\begin{thebibliography}{99}

\bibitem{Faddeev:1967}
L. D. Faddeev and V. N. Popov,
Phys. Lett. B {\bf 25}  (1967) 29.

\bibitem{Landau:1955}
L. D. Landau and I. M. Khalatnikov,
Zh. Eksp. Teor. Fiz. {\bf29}  (1955) 89 [Sov. Phys. JETP {\bf2} (1956) 69].

\bibitem{Fradkin:1955}
E. S. Fradkin,
Zh. Eksp. Teor. Fiz. {\bf29} (1955) 258 [Sov. Phys. JETP {\bf2} (1956) 361].

\bibitem{Johnson:1959}
K. Johnson and B. Zumino,
Phys. Rev. Lett. {\bf 3 } (1959)  351.

\bibitem{Zumino:1960}
B. Zumino,
J. Math. Phys. {\bf 1} (1960) 1.

\bibitem{Burden:1993}
C. J. Burden and C. D. Roberts,
Phys. Rev. D {\bf 47} (1993) 5581.

\bibitem{Bashir:1999}
A. Bashir, A. Kizilersu, M.R. Pennington,
ADP-99-8-T353, DTP-99-76, (1999).

\bibitem{Bashir:2000}
A. Bashir,
Phys. Lett. B {\bf 491} (2000) 280.

\bibitem{Bashir:2002}
A. Bashir and A. Raya,
Phys. Rev. D {\bf 66} (2002) 105005.

\bibitem{Delbourgo:2004}
A. Bashir and R. Delbourgo,
J. Phys. A {\bf 37} (2004) 6587.

\bibitem{Raya:2005}
A. Bashir and A. Raya,
Nucl. Phys. B {\bf 709} (2005) 307.

\bibitem{Bashir:2007}
A. Bashir and A. Raya,
Few Body Syst. {\bf 41} (2007) 185.

\bibitem{Bashir:2008}
A. Bashir, A. Raya and S. Sanchez-Madrigal,
J. Phys. A {\bf 41} (2008) 505401.

\bibitem{Bashir:2009}
A. Bashir, A. Raya, S. Sanchez-Madrigal and C. D. Roberts,
Few Body Syst. {\bf 46} (2009) 229.

\bibitem{Ahmadiniaz:2016}
N. Ahmadiniaz, A. Bashir and C. Schubert,
Phys. Rev. D {\bf 93} (2016) no.4, 045023.

\bibitem{Concha:2016}
A. Ahmad, J.J. Cobos-Martínez, Y. Concha-Sanchez and A. Raya,
Phys. Rev. D93 (2016) no.9, 094035.

\bibitem{Jia:2017pl}
S. Jia and M. R. Pennington,
Phys. Lett. B {\bf 769} (2017) 146.

\bibitem{Jia:2017pr}
S. Jia and M. R. Pennington,
Phys. Rev. D {\bf 95} (2017) 076007.

\bibitem{Dallolio:2019}
P. Dall'Olio and A. Bashir,
J. Phys. Conf. Ser. {\bf 1208} (2019) no.1, 012002.

\bibitem{Kotikov:2019}
A.V. Kotikov and S. Teber,
Phys. Rev. D {\bf 100} (2019) no.10, 105017.

\bibitem{Aslam:2016}
M. J. Aslam, A. Bashir and L. X. Guitierrez-Guerrero,
Phys. Rev. D {\bf93} (2016) no.7 076001.

\bibitem{Slavnov:1972}
A. A. Slavnov, Theor. Math. Phys. {\bf 10} (1972) 99;
J. C. Taylor, Nucl. Phys. B {\bf 33} (1971) 436.

\bibitem{Ward:1950}
J. Ward, Phys. Rev. {\bf 78} (1950) 182;
Y. Takahashi, Nuovo Cimento {\bf 6} (1957) 370.

\bibitem{Becchi:1976}
C.Becchi, A. Rouet and R. Stora,
Ann. Phys. {\bf 98} (1976) 287;
I. V. Tyutin, Lebedev Institute Preprint N39 (1975).

\bibitem{Curtis:1990}
D.C. Curtis, M.R. Pennington,
Phys. Rev. D {\bf 42} (1990) 4165.

\bibitem{Bashir:1994}
A. Bashir and M. R. Pennington,
Phys. Rev. D {\bf 50} (1994) 7679.

\bibitem{Delbourgo:2007}
A. Bashir, Y. Concha-Sanchez and R. Delbourgo,
Phys. Rev. D {\bf 76} (2007) 065009.

\bibitem{Kizilersu:2009}
A. Kizilersu, M.R. Pennington,
Phys.Rev. D {\bf 79} (2009) 125020.

\bibitem{Bashir:2011}
A. Bashir, A. Raya and S. Sanchez-Madrigal,
Phys. Rev. D {\bf 84} (2011) 036013.

\bibitem{Aguilar:2011}
A. C. Aguilar and J. Papavassiliou,
Phys. Rev. D {\bf 83} (2011) 014013.

\bibitem{Raya:2011}
A.~Bashir, A.~Raya and S.~Sanchez-Madrigal,
Phys. Rev. D {\bf 84} (2011) 036013.

\bibitem{Bashir:2012}
A. Bashir, R. Bermudez, L. Chang and C. D. Roberts,
Phys. Rev. C {\bf 85} (2012) 045205.

\bibitem{Qin:2013}
S.-X. Qin, Y.-X.~Liu, C.~D.~Roberts and S.~M.~Schmidt,
Phys. Lett. B {\bf 722} (2013) 384.

\bibitem{Kizilersu:2015}
A. K\i z\i lers\"{u}, T. Sizer, M. R. Pennington, A. G. Williams and R. Williams,
Phys. Rev. D {\bf 91} (2015) no. 6, 065015.
 	
\bibitem{Albino:2016}
L. Albino Fernandez-Rangel, A. Bashir, L.X. Gutierrez-Guerrero, Y. Concha-Sanchez,
Phys.Rev. D93 (2016) no.6, 065022.

\bibitem{Aguilar:2017}
A. C. Aguilar, J. C. Cardona, M. N. Ferreira and J. Papavassiliou,
Phys. Rev. D {\bf 96}  (2017) 014029.

\bibitem{Albino:2017}
R. Bermudez, L. Albino, L.X. Gutierrez-Guerrero, M.E. Tejeda-Yeomans, A. Bashir,
Phys.Rev. D95 (2017) no.3, 034041.

\bibitem{Albino:2019}
L. Albino, A. Bashir, L.X. Gutierrez Guerrero, B. El Bennich, E. Rojas,
Phys. Rev. D {\bf 100} (2019) no.5, 054028.

\bibitem{Roberts:1994}
C. D. Roberts and A. G. Williams,
Prog. Part. Nucl. Phys. {\bf 33} (1994) 477.

\bibitem{Alkofer:2001}
R. Alkofer and L. von Smekal,
Phys. Rept. {\bf 353} (2001) 281.

\bibitem{DeMeerleer:2018txc}
T.~De Meerleer, D.~Dudal, S.~P.~Sorella, P.~Dall'Olio and A.~Bashir,
Phys.\ Rev.\ D {\bf 97} no. 7, 074017 (2018)

\bibitem{Lavelle:1995ty}  M.~Lavelle and D.~McMullan,
Phys.\ Rept.\ \textbf{279} (1997) 1.

\bibitem{Gribov:1978}
V. N. Gribov,
Nucl. Phys. B {\bf 139} (1978) 1.

\bibitem{Capri:2015brs}
M. A. L. Capri \emph{et al.},
Phys. Rev. D {\bf 92} (2015) no. 4, 045039.

\bibitem{Capri:2016lin}
M. A. L. Capri \emph{et al.},
Phys. Rev. D {\bf 93} (2016) no. 6, 065019.

\bibitem{Capri:2016grib}
M. A. L. Capri \emph{et al.},
Phys. Rev. D {\bf 94} (2016) no. 2, 025035.

  \bibitem{Sonoda:2001}
H. Sonoda,
Phys. Lett. B {\bf 499} (2001) 253.

\bibitem{Capri:2016a2}
  M. A. L. Capri, D.~Fiorentini, M.~S.~Guimaraes, B.~W.~Mintz, L.~F.~Palhares and S.~P.~Sorella,
  Phys.\ Rev.\ D {\bf 94}  (2016) no. 6, 065009.

    \bibitem{Capri:2017ren}
  M. A. L. Capri, D. Fiorentini, A. D. Pereira and S. P. Sorella,
  Phys. Rev. D {\bf 96} (2017) no. 5, 054022.


  \bibitem{Nakanishi:1966}
  N. Nakanishi, Prog. Teor. Phys. {\bf 35} (1966) 1111;
  B. Lautrup, Kgl. Dan. Vid. Se. Mat. Fys. Medd. {\bf 35} (11) (1967) 1.

   \bibitem{Chetyrkin:1999}
  K. G. Chetyrkin, S. Narison and V. I. Zakharov,
  Nucl. Phys. B {\bf 550} (1999) 353.

  \bibitem{Gubarev:2001as}
  F. V. Gubarev, L. Stodolsky and V. I. Zakharov,
  Phys. Rev. Lett. {\bf 86} (2001) 2220.

  \bibitem{Gubarev:2001}
  F. V. Gubarev and V. I. Zakharov,
  Phys. Lett. B {\bf 501} (2001) 28. 

  \bibitem{Boucaud:2001}
  P. Boucaud, A. Le Yaouanc, J. P. Leroy, J. Micheli, O. Pene and J. Rodriguez-Quintero,
  Phys. Rev. D {\bf 63} (2001) 114003.

   \bibitem{Verschelde:2001}
  H. Verschelde, K. Knecht, K. Van Acoleyen and M. Vanderkelen,
  Phys. Lett. B {\bf 516} (2001) 307.

  \bibitem{Cornwall:1982}
 J. M. Cornwall,
 Phys. Rev. D {\bf 26} (1982) 11453.

 \bibitem{Aguilar:2004}
 A. C. Aguilar and A. A. Natale,
 JHEP {\bf 0408} (2004) 057.

  \bibitem{Aguilar:2008xm}
 A.~C.~Aguilar, D.~Binosi and J.~Papavassiliou,
  Phys.\ Rev.\ D {\bf 78} (2008) 025010.

 \bibitem{Dudal:2008}
 D. Dudal, J. A. Gracey, S. P. Sorella, N. Vandersickel and H. Verschelde,
 Phys. Rev. D {\bf 78} (2008) 065047.

\bibitem{Ayala:2012pb}
 A.~Ayala, A.~Bashir, D.~Binosi, M.~Cristoforetti and J.~Rodriguez-Quintero,
  Phys.\ Rev.\ D {\bf 86} (2012) 074512.

\bibitem{Siringo:2015wtx}
 F.~Siringo,
  Nucl.\ Phys.\ B {\bf 907} (2016) 572.

\bibitem{Cyrol:2017ewj}
 A.~K.~Cyrol, M.~Mitter, J.~M.~Pawlowski and N.~Strodthoff,
  Phys.\ Rev.\ D {\bf 97} (2018) no. 5, 054006.


   \bibitem{Cucchieri:2007}
  A. Cucchieri and T. Mendes,
  Proc. Sci., LAT2007 (2007) 297.

   \bibitem{Cucchieri:2008}
  A. Cucchieri and T. Mendes,
  Phys. Rev. Lett. {\bf 100} (2008) 241601.

  \bibitem{Bogolubsky:2007}
  I. L. Bogolubsky \emph{et al.},
  Proc. Sci., LAT2007 (2007) 290.

\bibitem{Oliveira:2012eh}
 O.~Oliveira and P.~J.~Silva,
  Phys.\ Rev.\ D {\bf 86} (2012) 114513.


   \bibitem{Tissier:2011}
  M. Tissier and N. Wschebor,
  Phys. Rev. D {\bf 84} (2011) 045018.

\bibitem{Gracey:2019xom}
 J.~A.~Gracey, M.~Peláez, U.~Reinosa and M.~Tissier,
  Phys.\ Rev.\ D {\bf 100} (2019) no. 3, 034023.


    \bibitem{Capri:2018ir}
  M. A. L. Capri \emph{et al.},
  Annals Phys. {\bf 390} (2018) 214.

   \bibitem{Ruegg:2004}
  H. Ruegg and M. Ruiz-Altaba,
  Int. J. Mod. Phys. A {\bf 19} (2004) 3265.

    \bibitem{Capri:2018un}
  M. A. L. Capri \emph{et al.},
  Phys. Lett. B {\bf 781} (2018) 48.

  \bibitem{Alloul:2014}
  A. Alloul \emph{et al.},
  Comput. Phys. Commun. {\bf 185} (2104) 2250.


   \bibitem{Hahn:2000}
  T. Hahn,
  Comput. Phys. Commun. {\bf 140} (2001) 418.

  \bibitem{Shtabovenko:2016}
  V. Shtabovenko, R. Mertig and F. Orellana,
  Comput. Phys. Commun. {\bf 207} (2016) 432;
  R. Mertig, M. B\"{o}hm and A. Denner,
  Comput. Phys. Commun. {\bf 64} (1991) 345.

   \bibitem{Grozin:2005}
  A. Grozin,
  ``Lectures on QED and QCD'', [hep-ph/0508242].

\bibitem{Nielsen:1975fs}
 N.~K.~Nielsen,
  Nucl.\ Phys.\ B {\bf 101} (1975) 173.

\bibitem{Gambino:1999ai}
 P.~Gambino and P.~A.~Grassi,
  Phys.\ Rev.\ D {\bf 62} (2000) 076002.
  [hep-ph/9907254].

  \bibitem{Capri:2016gut}
  M.~A.~L.~Capri, D.~Dudal, A.~D.~Pereira, D.~Fiorentini, M.~S.~Guimaraes, B.~W.~Mintz, L.~F.~Palhares and S.~P.~Sorella,
  Phys.\ Rev.\ D {\bf 95} (2017) no. 4, 045011.


\bibitem{Quadri:2014jha}
   A.~Quadri,
  Theor.\ Math.\ Phys.\  {\bf 182} no. 1, 74 (2015)
  [Teor.\ Mat.\ Fiz.\  {\bf 182} no. 1, 91 (2014)].


  \bibitem{Piguet:1984js}
  O.~Piguet and K.~Sibold,
  ``Gauge Independence in Ordinary {Yang-Mills} Theories,''
  Nucl.\ Phys.\ B {\bf 253} (1985) 517.

  \bibitem{Piguet:1995er}
  O.~Piguet and S.~P.~Sorella,
  ``Algebraic renormalization: Perturbative renormalization, symmetries and anomalies,''
  Lect.\ Notes Phys.\ M {\bf 28} (1995) 1.


   \bibitem{Fiorentini:2016rwx}
  M.~A.~L.~Capri, D.~Fiorentini, M.~S.~Guimaraes, B.~W.~Mintz, L.~F.~Palhares and S.~P.~Sorella,
  Phys.\ Rev.\ D {\bf 94} (2016) no. 6, 065009.








\end{thebibliography}
\end{document}